\documentclass[camera,letterpaper,nomarginnotes,nonarrowgutter]{jpaper}

\usepackage[normalem]{ulem}
\usepackage{mystyle}
\usepackage{xspace}
\usepackage{amsmath}
\usepackage{multirow}
\usepackage{array}
\usepackage{booktabs}
\usepackage{rotating}
\usepackage{balance}
\usepackage[normalem]{ulem}
\usepackage{colortbl}
\usepackage{listings}
\usepackage{setspace}
\usepackage{enumerate}
\usepackage{placeins}
\usepackage[hang,flushmargin]{footmisc}

\usepackage{fancyhdr}
\usepackage[compact]{titlesec}

\makeatletter
\g@addto@macro{\normalsize}{%
  \setlength{\abovedisplayskip}{2pt plus 1pt minus 1pt}
  \setlength{\belowdisplayskip}{2pt plus 1pt minus 1pt}
  \setlength{\abovedisplayshortskip}{0pt}
  \setlength{\belowdisplayshortskip}{0pt}
  \setlength{\intextsep}{2pt plus 1pt minus 1pt}
  \setlength{\textfloatsep}{4pt plus 1pt minus 1pt}
  \setlength{\skip\footins}{5pt plus 1pt minus 1pt}}
\makeatother

\usepackage{graphicx}
\usepackage{algpseudocode}
\usepackage{algorithm}
\usepackage{amsmath}

\usepackage{macros}
\usepackage{algpseudocode}

\usepackage{pifont}
\usepackage{color}
\usepackage{xcolor}
\usepackage{multirow}
\usepackage{stfloats}
\usepackage{listings}
\usepackage{caption}
\usepackage{subcaption}
\newcommand{\X}[0]{MetaSys\xspace}
\newcommand{\highlight}[1]{{#1}}
\newcommand{\revision}[1]{{#1}}
\newcommand{\rev}[1]{{#1}}
\newcommand{\tacorev}[1]{\textcolor{black}{#1}}
\definecolor{ceruleanblue}{rgb}{0.16, 0.32, 0.75}

\newcommand{\remove}[1]{}
\newcommand{\ignore}[1]{}
\newcommand{\thoughts}[1]{}
\newcommand{\One}{\emph{(i)}~}
\newcommand{\Two}{\emph{(ii)}~}
\newcommand{\Three}{\emph{(iii)}~}
\newcommand{\Four}{\emph{(iv)}~}

\newcommand{\Map}{\texttt{MAP}\xspace}

\newcommand{\TableMap}{\texttt{(UN)MAP}\xspace}
\newcommand{\TableMaptwo}{\texttt{(UN)MAP2D}\xspace}
\newcommand{\TableMapthree}{\texttt{(UN)MAP3D}\xspace}
\newcommand{\TableActivate}{\texttt{(DE)ACTIVATE}\xspace}

\newcommand{\Create}{\texttt{CREATE}\xspace}
\newcommand{\mysubsection}[1]{%
  \par
    \pagebreak[2]%
	  \refstepcounter{subsection}%
	      \everypar={%
		        {\setbox0=\lastbox}
				      \addcontentsline{toc}{subsection}{%
					          {\protect\makebox[0.3in][r]{\thesubsection.}
							  \hspace*{3pt}#1}}%
							        \textbf{\thesubsection\space\space{#1}. }%
									      \everypar={}%
										      }%
											    \ignorespaces
												}

\usepackage{url}

\begin{document}

\newcommand{\affilUT}[0]{\textsuperscript{*}}
\newcommand{\affilETH}[0]{\textsuperscript{\S}}
\newcommand{\affilCMU}[0]{\textsuperscript{$\dagger$}}

\title{\X: A Practical Open-Source Metadata Management System\\to Implement and Evaluate Cross-Layer Optimizations}

\author{
{Nandita Vijaykumar\affilUT}\qquad%
{Ataberk Olgun\affilETH}\qquad
{Konstantinos Kanellopoulos\affilETH}\qquad
{Hasan Hassan\affilETH}\qquad\\
{Mehrshad Lotfi\affilETH}\qquad%
{Phillip B. Gibbons\affilCMU}\qquad%
{Onur Mutlu\affilETH}\qquad\vspace{-3mm}\\\\
{\vspace{-3mm}\affilUT \emph{University of Toronto} \qquad \affilETH \emph{ETH Z{\"u}rich}} \qquad \affilCMU \emph{Carnegie Mellon University}
}

\maketitle

\thispagestyle{plain}
\pagestyle{plain}

\begin{abstract}
This paper introduces the first open-source FPGA-based infrastructure, \X, with a prototype
in a RISC-V {system}, to enable the rapid implementation and evaluation of a wide range of cross-layer techniques in real hardware. Hardware-software cooperative techniques are powerful approaches to improv{ing} the performance, quality of service, and security of general-purpose processors. They are however typically challenging to rapidly implement and evaluate in real hardware as they require full-stack changes to the hardware, system software, and instruction-set architecture (ISA).

\X implements a rich hardware-software interface and lightweight metadata support that can be used as a common basis to rapidly implement and evaluate new cross-layer techniques. We demonstrate \X's versatility and ease-of-use by implementing and
evaluating three cross-layer techniques for: \One prefetching {in} graph analytics; \Two bounds checking in memory unsafe languages, and \Three return address protection in stack frames; each technique requiring {only} {\textasciitilde}100 lines of Chisel code over MetaSys.

Using \X, we perform the first detailed experimental study to quantify the performance overheads of using a \emph{single} metadata management system to enable multiple cross-layer optimizations in CPUs. We identify the key sources of bottlenecks and system inefficiency of a general metadata management system. We design MetaSys to minimize these inefficiencies and provide increased versatility compared to previously-proposed metadata systems. Using three use cases and a detailed characterization, we demonstrate that a common metadata management system can be used to efficiently support diverse cross-layer techniques in CPUs. {MetaSys is completely and freely available at \url{https://github.com/CMU-SAFARI/MetaSys}}

\end{abstract}

\section{Introduction}

Hardware-software cooperative techniques offer a powerful approach to {improving} the performance and efficiency of general-purpose processors. These techniques involve communicating key application {and semantic} information from the software to the architecture to enable more powerful optimizations and resource management in hardware. 
Recent research proposes many such cross-layer
{techniques} for various purposes, e.g., performance, quality of service (QoS), memory protection, programmability, security. For example, Whirlpool~\cite{whirlpool-mukkara-asplos16} identifies and communicates 
regions of memory that have similar properties (i.e., data structures) in
the program to the
hardware, {which uses this information} to more intelligently place data in a
non-uniform cache architecture (NUCA) system.
RADAR~\cite{radar-manivannan-hpca16} and
EvictMe~\cite{evictme-wang-pact02} communicate
which {cache} blocks will no longer be used in the program, such that cache 
policies can evict them. 
These are just a few examples in an increasingly large space of cross-layer
techniques proposed in the form of hints implemented as new ISA
instructions to aid cache replacement, prefetching, {memory management,} etc.~\cite{swcache-jain-iccad01,compilerpartitioned-ravindran-lctes07,popt-gu-lcpc08,pacman-brock-ismm13,evictme-wang-pact02,generatinghints-beyls-jsysarch05,keepme-sartor-interact05,compilerassisted-yang-lcpc04,cooperativescrubbing-sartor-pact14,
runtimellc-pan-sc15,prefetchtasklifetimes-papaefstathiou-ics13,radar-manivannan-hpca16,modified-tyson-micro95}, 
program annotations/directives to convey program
semantics~\cite{pageplacement-agarwal-asplos15,whirlpool-mukkara-asplos16,datatiering-dulloor-eurosys16,
flikker-liu-asplos11,popt-gu-lcpc08,ldesc}, or interfaces to communicate an application's QoS requirements for efficient partitioning and prioritization of shared hardware resources~\cite{pard-ma-asplos15,labeled-yu}.

While cross-layer approaches have been demonstrated to be highly effective, such proposals are challenging to evaluate on real hardware as they require cross-layer changes to the hardware, operating system (OS), {application software, and} instruction-set architecture (ISA). Existing open-source infrastructure{s} for implementing cross-layer techniques in real hardware include PARD~\cite{labeled-yu,pard-ma-asplos15} for QoS and Cheri~\cite{cheri} for fine-grained memory protection and security. Unfortunately, these \revision{infrastructures} are not designed to provide key features required for \emph{performance} optimizations: \One rich dynamic hardware-software interfaces, \Two low-overhead metadata management, and \Three interfaces to numerous hardware components such as prefetchers, caches, etc.

In this work, we introduce \X (\textbf{Meta}data Management \textbf{Sys}tem for Cross-Layer Performance Optimization), a full-system FPGA-based infrastructure, with a prototype in the RISC-V
{Rocket Chip system}~\cite{rocket-chip-gen}, to enable rapid implementation and evaluation of diverse cross-layer techniques in real hardware. 
\X comprises three key components:  
(1) A rich \textbf{hardware-software interface} to communicate a \emph{general} and extensible set of application information to the hardware architecture at runtime. \tacorev{We refer to this additional application information as \emph{metadata}. {Examples of metadata} include memory access pattern information for prefetching, data reuse information for cache management, address bounds for hardware bounds checking, etc.} The interface is implemented as new instructions in the RISC-V ISA and is wrapped with easy-to-use software library abstractions. 
(2) \textbf{Metadata management} support in the OS and hardware to {store and access} the communicated metadata. Hardware components performing optimizations can then efficiently query for the metadata. We use a \emph{tagged memory-based} design for metadata management where each memory address is tagged with an ID. This ID points to metadata that describes the data contained in the {location specified by the} memory address. (3)~\textbf{Modularized components} to quickly implement various cross-layer optimizations with interfaces to the metadata management support, OS, core, and memory system. {Our FPGA-based infrastructure provides flexible modules that can be easily extended to implement different cross-layer optimizations.}

{The closest work to our proposed system is XMem~\cite{xmem}.} XMem proposes a general metadata management system that {can communicate semantic information at compile time. This limits the use cases supported by XMem.} \tacorev{\X has the following benefits over XMem:} First, \X offers a richer interface that communicates a flexible amount of metadata at \emph{runtime}, rather than being limited to statically available program information. This enables a wider set of use cases and more powerful cross-layer techniques (as explained in §\ref{sec:xmem}).
Second, \X has a more optimized system design that is designed to be \emph{lightweight} in terms of the hardware complexity and changes to the ISA, without sacrificing versatility (§\ref{sec:xmem}). \X incurs only a small area overhead of 0.02\% (including 17KB of additional SRAM), \revision{0.2\% memory overhead in DRAM}, and adds only 8 new instructions to the RISC-V ISA. {Third, MetaSys is open-source and freely available, whereas XMem} {is neither} implemented nor evaluated in real hardware with full-system support.

\textbf{Use cases.} Cross-layer techniques that can be implemented with \X include performance optimizations such as cache management, prefetching,
memory scheduling, data compression, and data placement; cross-layer techniques for QoS; and lightweight techniques for memory protection (see §\ref{sec:use-cases}). 
To demonstrate the versatility and ease-of-use of \X in implementing new cross-layer techniques, we implement and evaluate three hardware-software cooperative techniques: \One prefetching for graph analytics applications; \Two bounds checking in memory unsafe languages, and \Three return address protection in stack frames. These techniques were quick to implement with \X, each requiring {only} an additional {\textasciitilde}100 lines of Chisel~\cite{chisel} code {on top of MetaSys's hardware codebase ({\textasciitilde}1800 lines of code)}. 

\textbf{Characterizing a general metadata management system.} 
Using \X, we perform the first detailed
experimental characterization and limit study of the performance overheads of using a \emph{single} common metadata management system to enable multiple diverse cross-layer techniques in a general-purpose processor. We make {four} new observations from our characterization across 24 applications and 4 microbenchmarks that were designed to stress \X.

First, the performance overheads from the cross-layer interface and metadata system itself are on average very low (2.7\% on average, up to 27\% for the most intensive microbenchmark).
Second, there is no performance loss from supporting \emph{multiple} techniques that simultaneously query the shared metadata system. This indicates that {\X{}} can be designed to be a scalable substrate. Third, the most critical factor in determining the performance overhead is the fundamental spatial and temporal locality in the accesses to the metadata itself. This determines the effectiveness of the metadata caches and the additional memory accesses to retrieve metadata. {Fourth, we identify TLB misses} from the required address translation when metadata is retrieved from memory as an important factor in performance overhead.

\textbf{Conclusions from characterization.} From our detailed characterization and implemented use cases on real hardware, we make the following conclusions: {First, }using a \emph{single} general metadata management system is a promising low-overhead approach to implement \emph{multiple} cross-layer techniques in future general-purpose processors. The significance of using a single framework is in enabling a wide range of cross-layer techniques with a single change to the hardware-software interface~\cite{xmem,pard-ma-asplos15} and \emph{consolidating} common metadata management support; thus, making the adoption of new cross-layer techniques in future processors significantly easier. {Second,} we demonstrate that a common framework can simultaneously and scalably support multiple {cross-layer} optimizations. For our implemented use cases, we observe low performance overheads from using the general \X system: 0.2\% for prefetching, 14\% for bounds checking, and 1.2\% for return address protection.

{\X{} is fully open-source and freely available at \url{https://www.github.com/CMU-SAFARI/MetaSys}.}

This work makes the following major contributions.
\begin{itemize}
    \item We introduce \X, the first full-system open-source FPGA-based infrastructure of a lightweight metadata management system{. \X{} provides a} rich hardware-software interface that can be used to implement a diverse set of cross-layer techniques. {We implement a prototype of \X in a RISC-V system providing} the required support in the hardware, OS, and the ISA to enable quick implementation and evaluation of new hardware-software cooperative techniques in real hardware.
    \item \tacorev{We propose a new hardware-software interface that enables \emph{dynamically} communicating information and a more streamlined system design that can support a richer set of cross-layer optimizations than prior work{~\cite{xmem}}.}
        \item We present the first detailed experimental characterization of the performance and area overheads of a \emph{general} hardware-software interface and lightweight metadata management system designed to enable \emph{multiple} and diverse cross-layer performance optimizations. We identify key sources of inefficiencies and bottlenecks of a general metadata system on real hardware, and we demonstrate its effectiveness as a common substrate for enabling cross-layer techniques in CPUs.
     \item We demonstrate the versatility and ease-of-use of the \X infrastructure by implementing and evaluating three hardware-software cooperative techniques: \One prefetching for graph analytics applications; \Two efficient bounds checking for memory-unsafe languages; and \Three return address protection for stack frames. We highlight other use cases that can be implemented with \X.   

\end{itemize}

\section{Background and Related Work}

\textbf{Hardware-software cooperative techniques in
CPUs.} 
Cross-layer performance optimizations communicate \emph{additional}
information across the application-system boundary{. W}e refer to this information as \emph{metadata}. Metadata that is typically useful for performance optimization include program properties such as access patterns, read-write
characteristics, data locality/reuse, data types/layouts, data
"hotness", and working set size. This metadata enables
more intelligent hardware/system optimizations such as cache management, data placement, thread scheduling,
memory scheduling, data compression, and approximation~\cite{xmem,ldesc,nanditaThesis}. For QoS
optimizations, metadata includes application priorities and prioritization rules for
allocation of resources such as memory bandwidth and cache
space~\cite{pard-ma-asplos15, labeled-yu, parbs-mutlu-isca08, mutlu2007stall,tcm-kim-micro10,ebrahimi-10,subramanian2013mise,subramanian2015asm}. Memory safety optimizations may communicate base/bounds addresses of data structures~\cite{pump,hardbound}.

A \emph{general} framework is a promising approach as it enables many cross-layer techniques with a single change to the hardware-software interface and enables \emph{reusing} the metadata management support across multiple optimizations. Such systems were recently proposed for performance~\cite{xmem,ldesc}, memory protection and security~\cite{pump,cheri}, and QoS~\cite{pard-ma-asplos15,labeled-yu}. 

A general framework to support a wide range of cross-layer optimizations{\textemdash}specifically for \emph{performance}{\textemdash}requires: \One a rich and dynamic hardware-software interface to communicate a diverse set of metadata at runtime and \Two lightweight and \emph{low-overhead} metadata management~\cite{xmem}{, and \Three interfaces to numerous hardware components}. Even small overheads imposed as a result of the system's generality may overshadow the performance benefits of a cross-layer technique. General metadata systems may also impose
significant complexity, performance, and power overheads to the processor. 
While prior work has demonstrated the significant benefits of cross-layer approaches, no
previous work has characterized the efficiency and capacity limits of a general metadata system for cross-layer optimizations in CPUs.

\textbf{Tagged architectures.} \X is inspired by the metadata management and interfaces proposed in XMem~\cite{xmem} and the large body of work on tagged memory~\cite{mondrian-witchel-asplos02,efficient-tagged-memory,pump,tagged-feustel-taco73,hardware-zeldovich-osdi08} and capability-based systems~\cite{cheri,carter1994hardware,levy2014capability,watsoncapability}. We compare against the closest {prior} work, XMem, {qualitatively} in §\ref{sec:xmem} and quantitatively in §\ref{sec:prefetching}. Unlike all above works, our goal is to provide an open-source framework to implement and {these prior} cross-layer approaches in real hardware and to perform a {detailed real-system} characterization of such metadata systems for performance optimization. 

\textbf{Infrastructure for evaluating cross-layer techniques.}
Evaluating the overheads and feasibility of a newly-proposed
cross-layer technique is non-trivial. Fully characterizing the performance and
area overheads either with a full-system
cycle-accurate simulator or an FPGA implementation requires implementing: \One \emph{Hardware
support} to implement the mechanism; \Two
\emph{OS support} for OS-based cross-layer optimizations and to characterize the context-switch and system
overheads of saving and handling a process' metadata; and
\Three \emph{Compiler support and ISA modifications} to add and recognize new instructions to
communicate metadata. 

Recent works propose general systems that are designed to enable cross-layer techniques for QoS (PARD~\cite{labeled-yu,pard-ma-asplos15}) or fine-grained memory protection and security (Cheri~\cite{cheri}). 
 PARD enables tagging of components and applications with IDs that are propagated with memory requests and enforc{ing} QoS requirements in hardware. Cheri~\cite{cheri} is a capability-based system that provides hardware support and ISA extensions to enable fine-grained memory protection. Neither system supports the \One communication of diverse metadata at runtime, \Two flexible granularity tagging of memory to enable efficient metadata lookups from multiple components, or \Three  interfaces to numerous hardware components {(such as the prefetcher, caches, memory controllers) that are} needed for \emph{performance} optimization.

\underline{\textbf{Our Goal.}}
Our goal in this work is twofold. 
First, we aim to develop an efficient and flexible \emph{open-source} framework that enables rapid implementation of new cross-layer techniques to evaluate the associated performance, area, and power overheads, and thus their {benefits and} feasibility, in real hardware.  

Second, we aim to perform the first detailed limit study to characterize and experimentally quantify the overheads associated with \emph{general} metadata systems to determine their practicality for performance optimization in future CPUs.

\section{\X: Enabling and {E}valuating {C}ross-layer {O}ptimizations} 
\label{sec:infra}
To this end, we develop \X (\textbf{Meta}data Management \textbf{Sys}tem for Cross-Layer Performance Optimization), an open-source full-system FPGA-based infrastructure to
implement and evaluate new cross-layer
techniques in real hardware. \X includes: \One a rich hardware-software interface to dynamically communicate a flexible amount of metadata at runtime from the application to the hardware, using new RISC-V instructions;
\Two a tagged memory-based~\cite{mondrian-witchel-asplos02,efficient-tagged-memory,pump,tagged-feustel-taco73,hardware-zeldovich-osdi08} implementation of metadata management in the system and OS; and \Three flexible modules to add new hardware optimizations with interfaces to the metadata, {processor}, memory, and OS.
We build a prototype of \X in the RISC-V Rocket {Chip~\cite{rocket-chip-gen} system}.  

We choose an FGPA implementation as opposed to a full-system simulator as: \One This enables {us to} focus on feasibility as all components need to be fully implemented (e.g., ports, wires, buffers) and {their} impact on area, cycle time, power, and scalability is quickly visible. \Two FPGAs are much faster, running full application simulations in a few minutes/hours as opposed to many days on a full-system simulator, making {FPGAs} a better fit for quick experimentation. \Three The RTL generated can be used for more accurate area and power calculation {and potential future synthesis on other systems}.

\begin{figure}[h]
\includegraphics[width=1.0\linewidth]{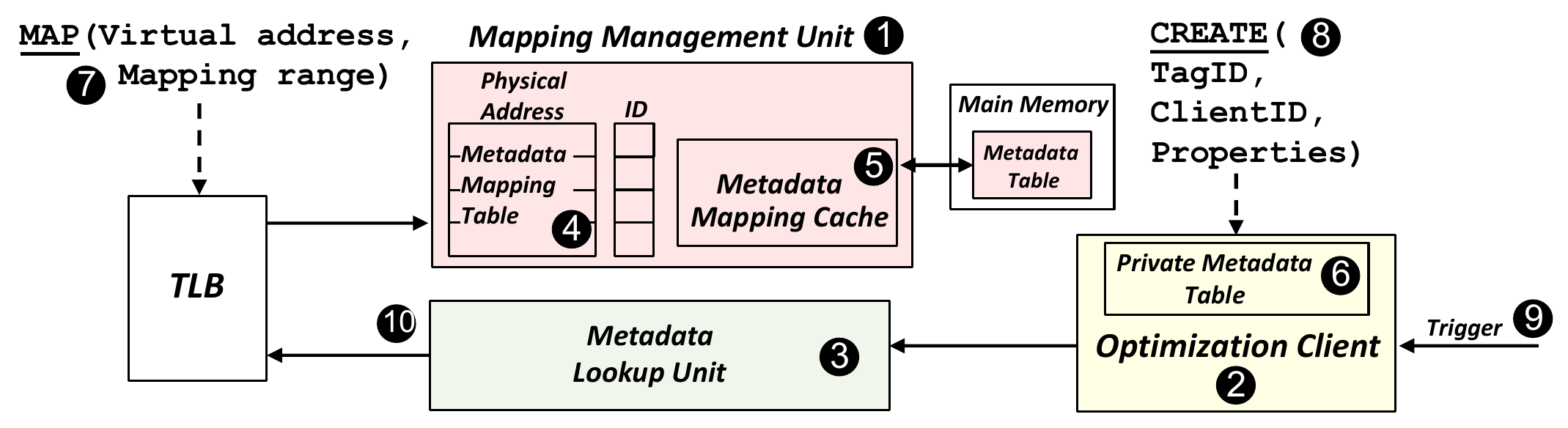}
\caption{\revision{\X hardware components and operation. {MetaSys' structures are highlighted.}}}
\label{fig:meta_interface}
\end{figure}

Fig.~\ref{fig:meta_interface} depicts an
overview of the major hardware components in \X and their operation: {The mapping management unit~\ding{182}, the optimization {client}~\ding{183}, and the metadata lookup unit \ding{184}.}

\subsection{Tagged {M}emory-based {M}etadata {M}anagement} 
Similar to prior systems for taint-tracking, security, and performance optimization, \X implements \emph{tagged memory}-based~\cite{mondrian-witchel-asplos02,efficient-tagged-memory,tagged-feustel-taco73,hardware-zeldovich-osdi08} metadata management. \X associates metadata with memory address ranges of {arbitrary sizes} by tagging each memory address with an 8-bit (configurable) ID or tag. Each tag is a
unique pointer to metadata that describes the data at the memory address. Hardware optimizations (e.g., in the cache, memory controller, or core) can query for the tag
associated with any memory address and the metadata associated with the tag.

\revision{The mapping between each memory address and the corresponding ID is saved in a
table in memory referred to as \highlight{Metadata Mapping Table} (MMT): {\ding{185}} in Fig.~\ref{fig:meta_interface}. This table is allocated by the OS for each process and is saved in memory. In \X (similar to XMem~\cite{xmem} and Cheri~\cite{cheri}), we
tag \emph{physical addresses}. As a result, any virtual address has to be
translated before indexing the \highlight{MMT} to retrieve the tag ID. To enable fast retrieval of
IDs, we implement a cache for the \highlight{MMT} in hardware that stores
frequently accessed mappings, referred to as the Metadata Mapping Cache (MMC)
{\ding{186}}. MMC misses lead to memory accesses to retrieve mappings from the MMT in memory.}

\X can be configured to tag memory at flexible
granularities. In §\ref{sec:analysis}, we evaluate the performance impact of the tagging
granularity. \revision{The size of the MMT depends on the tagging granularity. For a 512B mapping granularity, the MMT requires 0.2\% of physical memory (16MB in a 8GB system). The MMC holds 128 entries, where each entry stores a physical-address-to-tag mapping, and is 608B in size (8 bit entry and 30 bit tag).}

\tacorev{We implement dedicated mapping tables for tag IDs rather than use the page table or TLBs for the following reasons: First, {doing so} obviates the need to modify the latency{-}critical address translation structures. Second, MetaSys associates physical addresses with Tag IDs rather than virtual addresses (to enable the memory controller and LLCs to look up metadata). Thus, a page table or TLB cannot be directly used to save Tag IDs as they are indexed with virtual addresses.} 

The actual metadata associated with any ID is saved in special SRAM caches that
are private to each hardware component or optimization. For example,
the prefetcher would separately save access pattern information, while a
hardware bounds checker would privately save {data structure} bound{ary} information. We refer to
these stores as Private Metadata Tables (PMTs)~{\ding{187}}. \rev{The PMTs are saved near each component (private to each component) and are loaded/updated by MetaSys. The metadata (e.g., locality/”hotness”) is encoded such that it can be directly interpreted by the component, e.g., {a} prefetcher.}

\subsection{The Hardware-Software Interface} 
Communicating application information with \X requires \One associating memory
address ranges with a tag or ID of configurable size (8 bits by default) and \Two associating each ID with the
relevant metadata. The metadata could include
program properties that describe the memory range, such as data locality/reuse, access patterns, read-write
characteristics, data "hotness", and data types/layouts. We use two operators \revision{(described below)} that can be called in programs to dynamically communicate metadata.

To associate memory address ranges with an ID, we provide the
\texttt{MAP}/\texttt{UNMAP} 
interface {\ding{188}} (similar to XMem~\cite{xmem}). \Map and \texttt{UNMAP} are implemented as new RISC-V instructions
that are interpreted
by the {Mapping Management Unit (MMU)} to map a range of memory addresses (from a given
virtual address up to a certain length) to the provided
ID. \revision{These mappings are saved by the {MMU} in the MMT}. We also implement 2D and 3D versions of \Map to efficiently map
2/3-dimensional address ranges in a multi-dimensional data structure with a single instruction.

To associate each ID with metadata, we provide the \texttt{CREATE} interface.
\Create{} {\ding{189}} takes 3 inputs from the application: the tag ID,
the 8-bit ID for the
\emph{hardware component} (i.e., prefetcher, bounds checker, etc., \revision{called Module ID}), and 512B of
metadata. \tacorev{\Create directly populates the PMT of the
appropriate hardware component with 512 bytes of {(or less)} metadata. Each PMT (private to the optimization {client}) has 256 entries assuming {8-}bit tag IDs. The \Create operator overwrites the metadata at the entry indexed by the tag ID at the PMT specified by the module ID.} \rev{All \Create and \Map instructions are associated with the \emph{next} load/store instruction in program order to avoid inaccuracies due to out-of-order execution. In other words, an implicit dependence is created in hardware between these instructions and the next load/store, and they are committed together.} \tacorev{This enables associating information with the next load/store and not just the memory region associated with it, e.g., in the bounds checking use case described in §\ref{sec:boundschecking}.}

Table~\ref{table:semantics} lists the new instructions along with their arguments.

\begin{table}[h]
    \caption{\highlight{MetaSys instructions.}}
    \centering
    \scriptsize

    \begin{tabular}{@{} lm{25em} @{}}
\toprule
\textbf{MetaSys Operator} & \textbf{MetaSys ISA Instructions}\\        
\midrule
\Create & \Create {Client}ID, \revision{TagID}, Metadata \\
\midrule
\multirow{3}{5em}{\TableMap} & \TableMap TagID, start\_addr, size \\
& \TableMaptwo TagID, start\_addr, lenX, sizeX, sizeY \\
& \TableMapthree TagID, start\_addr, lenX, lenY, sizeX, sizeY, sizeZ ;  \\
\midrule
\label{table:semantics}
\end{tabular}
\end{table}

\subsection{Metadata Lookup}

Each optimization component is triggered by a hardware event \ding{190} (e.g., a cache
miss). \revision{{A component} then retrieves the physical address {corresponding to the virtual address associated with the event (e.g., the virtual address that misses in the cache)} from the TLB \ding{191} \tacorev{(in case of L1 optimizations)} and queries the MMC with the physical 
address to retrieve the associated tag ID.} On a
miss in the MMC, the mapping is retrieved from the MMT in memory. The
optimization {client} uses the retrieved tag ID to \revision{obtain} the appropriate
metadata \revision{from} the PMT. \tacorev{The optimization {client} is designed to flexibly implement a wide range of use cases and can be designed based on the optimization {at hand}. For example, {the optimization client used} to build the prefetcher use case in §\ref{sec:prefetching} {has} interfaces to the prefetcher, caches, memory controller, and TLBs to make implementing optimizations easier. Each {client} has a static ID ({client}ID) and a PMT that is updated by the \Create operator.}  

\subsection{Operating System Support}

We add OS support for metadata management in the RISC-V proxy
kernel~\cite{riscv-pk}, which can
be booted on our Rocket RISC-V prototype: \tacorev{First, we add support to manage the
MMT in memory, where the OS allocates the MMT in the physical address space and communicates the pointer to the MAP hardware support. Second, we add support to flush the \revision{PMTs} during a context
switch (similar to how the TLB is flushed). \revision{Third, if the OS changes the virtual to physical address mapping of a page, then to ensure consistency of the metadata, the MMT is updated by the OS to reflect the correct physical-address-to-tag-ID mapping and the corresponding MMC entries are invalidated. We modify the page allocation {mechanism in the} OS to do this.} In addition, we also provide support to implement optimizations performed by the
OS or with OS cooperation. {To do so,} \X enables trapping into the OS to perform customized
checks or optimizations (e.g., protection checks or altering virtual-to-physical
mappings) based on specific hardware trigger events (using interrupt routines).} We describe one such use
case in §\ref{sec:safety}. 

\subsection{\highlight{Coherence/Consistency of Metadata in Multicore {Systems}}}

\revision{MetaSys can be flexibly extended to multicore processors. \rev{Metadata is maintained at a process-level, therefore, threads within the same process cannot have different metadata for the same data structure.} The MMC is a per-core structure, while the Private Metadata Tables (PMTs) are per-component structures (e.g, at the memory controller, LLC, prefetcher). 
The two dynamic operators (\Create and \Map) may cause challenges in coherence and consistency of metadata in multicore systems. \Create directly updates metadata associated with the \emph{per-process} tag ID, which is saved at the per-component PMTs. \rev{The PMTs are shared by all cores when the optimization component is also shared (and thus any updates by \Create are automatically coherent). The PMTs for private components (e.g., L1 cache) are not coherent and can only be updated by the corresponding thread.} \Map updates the mapping in the MMC, which is private to each core. To ensure coherence of the MMC mappings, a \Map update invalidates the corresponding MMC entry (if present) in other MMCs \tacorev{by broadcasting updates with a snoop{y} protocol.} If the use case requires consistency of the metadata, i.e., ordering between a \Create/\Map instruction and when it is visible to other cores, barriers and fence instructions are used to enforce any required ordering between threads for updates to metadata.}

\subsection{Timing Sensitivity of Metadata}
\X supports three modes: \One \emph{Force stall}, where the instruction triggering a metadata lookup cannot commit \revision{until} the optimization completes (e.g., for security use cases); \Two \emph{No stall}, where metadata lookups do not stall the core but are always resolved (e.g., for page placement, cache replacement), and \Three \emph{Best effort}, where lookups may be dropped to minimize performance overheads (e.g., for prefetcher training). 

\subsection{\highlight{Software Library}}
\highlight{We develop a software library {that} can be included in user programs to facilitate the use
of MetaSys primitives \Create and \Map (Table~\ref{table:library}). The library exposes three functions: \One \Create 
populates an entry indexed by the tag ID (\textit{TagID}) in the PMT of a hardware optimization {client} (\textit{{Client}ID}) with the corresponding metadata;
\Two \texttt{MAP} updates the MMT by assigning tag IDs to memory addresses of the range (\textit{start, end}); \Three \texttt{UNMAP} resets the tag IDs of the corresponding address range in the MMT.} \tacorev{While the operators can be directly used via the provided software library, their use can be simplified by using wrapper libraries that abstract away the need to directly manage tag IDs and their mappings.}

\begin{table}[h]
    \centering
    \caption{\highlight{The MetaSys software library {function calls}.}}
    \scriptsize
    \begin{tabular}{@{} ll @{}}
    \toprule
    \textbf{Library {Function} Call} & \textbf{Description}\\        
    \midrule
	CREATE(\revision{\textit{{Client}ID}}, \textit{TagID}, \textit{*meta}) & \revision{\textit{{Client}ID}} -> PMT[\textit{TagID}] = \textit{*metadata}\\
    \midrule
    MAP(\textit{start*}, \textit{end*}, \textit{TagID}) & MMT[\textit{start...end}] = \textit{TagID}\\
    \midrule
    UNMAP(\textit{start*}, \textit{end*}) & MMT[\textit{start...end}] = 0\\
    \midrule
    \end{tabular}
    \label{table:library}
\end{table}

\subsection{Comparison to the XMem {F}ramework~\cite{xmem}}
\label{sec:xmem}
MetaSys implements a tagged-memory-based system with a metadata cache similar to XMem~\cite{xmem}. \X however has three major benefits {over XMem}. First, \X enables communicating metadata at \emph{runtime} using a more powerful \Create operator that is implemented as a new instruction. In XMem, metadata is communicated only \emph{statically} at compile time (\Create is hence a compiler pragma). \X thus enables a wider set of optimizations including {fine-grained} memory safety, protection, prefetching, etc., and enables communicating metadata that is dependent {on program input} and metadata that can be {accurately} known {only} at runtime (e.g., access patterns, data "hotness", etc.). \X was designed to efficiently handle these dynamic metadata updates. Second, the dynamic and more expressive \Create operator obviates the need for additional interfaces (\texttt{ACTIVATE}/\texttt{DEACTIVATE}) to track the validity of statically communicated metadata. This enables a more streamlined metadata system in \X with fewer {new instructions,} tables, and lookups. Third, MetaSys allows the application programmer to directly select which cross-layer optimization to enable/disable and communicate metadata to, via the \Create operator. XMem, on the other hand, does not allow control of hardware optimizations from the application. Table~\ref{table:operators} summarizes the {\X{}} operators and compares to the corresponding operators in XMem. \emph{Of the three {\X{}} use cases {we evaluate in this paper}, only return address protection (§\ref{sec:rap}) can be implemented with XMem.}
\begin{table*}[!t]
    \centering
    \caption{\highlight{Comparison between MetaSys and XMem interfaces.}}
        \label{table:operators}
\scriptsize
    \begin{tabular}{@{} lm{28em}m{34em} @{}}
    \toprule
    \textbf{Operator} & \textbf{XMem~\cite{xmem}} & \textbf{MetaSys}\\        
    \midrule
	\Create & Compiler pragma to communicate static metadata at {program} load time. & Selects a hardware optimization, dynamically associates metadata with an ID, and communicates both to hardware at runtime (implemented as {a} new instruction).\\
    \midrule
    \TableMap & Associate memory ranges with tag IDs (implemented as new instructions). & Same semantics and implementation as XMem.\\
    \midrule
    \TableActivate & Enable/disable optimizations associated with a tag ID (implemented as new instructions). & {\textbf{{Does not exist}}} as the same functionality can now be done with CREATE.\\
    \midrule
    \end{tabular}
\end{table*}

\subsection{FPGA-based {I}nfrastructure}
We build a full system prototype of \X on an FPGA with the Rocket {Chip RISC-V system}~\cite{rocket-chip-gen} and add the necessary support in the compiler, libraries, OS, ISA, and hardware. The modularized \X components can also be ported to other RISC-V cores. We used the RoCC accelerator~\cite{rocket-chip-gen} in the Rocket chip to implement
the metadata management system. RoCC is a customizable module that
enables interfacing with the core and memory. The hardware support implemented in ROCC comprises \One
the control logic to handle {\Map}s and {\Create}s, \Two control logic to perform
metadata lookups by components that implement optimizations, and \Three the {memory for} metadata
caches (MMC and PMTs). We extended the RISC-V ISA with 8 instructions (6 for \Map/\Create and 2 for OS operations). \highlight{To implement all the hardware modules of \X, we modified/added 1781 lines of Chisel code in the Rocket Chip. As we demonstrate later, since the \X hardware modules can be flexibly reused across multiple hardware-software optimizations, the techniques in our use cases only required 87-103 additional lines of Chisel code.} \tacorev{The full MetaSys infrastructure {is} open-sourced{~\cite{metasysGithub}} including the Chisel code for the \X hardware support, the RISC-V OS with the required modification, and the software libraries to expose the \X primitives.}

\subsection{Implementing a {H}ardware-{S}oftware {C}ooperative {T}echnique with \X}
\begin{figure}[h]
\includegraphics[width=1.0\linewidth]{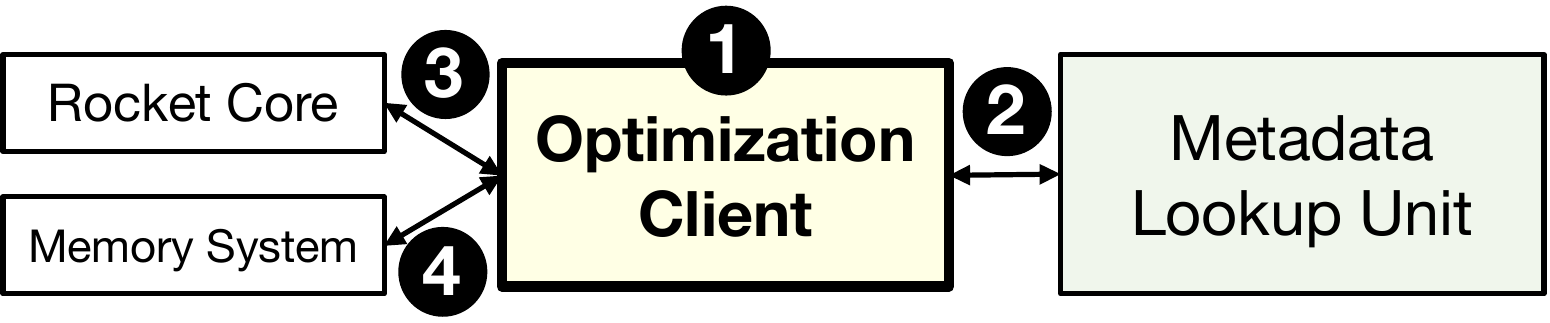}
\caption{\X Optimization {Client}.}
\label{fig:meta_module}
\end{figure}

To implement a new hardware technique with the baseline \X code, we provide a flexible module (\ding{182} in Fig.~\ref{fig:meta_module}) \rev{with a PMT and interfaces to the metadata lookup unit, to the core (to receive triggers), and interfaces to the cache controller.} The interface to the lookup unit \ding{183} provides dynamic access to the metadata communicated by the \Create and \Map operators. The interfaces to the core \ding{184} and the memory system \ding{185} can be used as \emph{trigger} events for optimization and lookups (e.g., a cache miss). The different components within the \X logic itself (i.e., the metadata caches, logic to access the Metadata Mapping Table in memory, and the lookup logic) can be flexibly reconfigured. 

\subsection{Dynamically-{T}yped or {M}anaged {L}anguages}
MetaSys relies heavily on function calls/libraries that abstract away low level details that call the \X instructions even in C/C++. With managed and dynamically-typed languages, the metadata associated with data structures/objects would be provided by the user with additional class/object member functions. The metadata could also be directly embedded within object/class definitions (e.g., a list or map in Python would by definition have certain access properties). Other properties (e.g., data types) would be provided by the interpreter (in the case of dynamically-typed languages) and the mapping/remapping calls to memory addresses would be handled by the runtime during memory (de)allocation.

\tacorev{\subsection{Comparison to {S}pecialized {C}ross-{L}ayer {S}olutions}
\vspace{-1.5mm}
In comparison to specialized cross-layer solutions, MetaSys offers the following benefits: (i) Generality: towards implementing a large number of use cases, including more complex use cases such as specialized prefetching (§\ref{sec:prefetching}), which amortizes the overall hardware cost; (ii) Flexibility and versatility in the implemented instructions: A challenge with specialized cross-layer solutions is the need to add new instructions that create challenges in forward/backward compatibility and also require changes across the stack for each new optimization. With MetaSys, the instructions are designed to be agnostic to the optimization and only require {a} one-off change to the hardware-software interface; (iii) Infrastructure for evaluation: MetaSys can be used to implement many specialized cross-layer techniques in real hardware, which would otherwise be a challenging programming task (as demonstrated in §\ref{sec:prefetching} and §\ref{sec:safety}). In §\ref{sec:prefetching} and §\ref{sec:safety}, we evaluate MetaSys's ability to implement several cross-layer techniques.
}

\section{Methodology}
\label{sec:methodology} 

\textbf{Baseline system.} We use the in-order Rocket core~\cite{rocket-chip-gen}
as the baseline CPU and conduct our experiments on the ZedBoard
Zynq-7000~\cite{zedboard} FPGA board.  
Table \ref{table:rocket_params} lists the parameters of the core and memory system {as well as evaluated workloads}.\footnote{\revision{Since DRAM is disproportionately faster than the CPU clock rate on FPGAs, we added logic in the memory controller to scale the rate at which memory requests are issued. The resulting average memory latency and bandwidth in core cycles were validated with microbenchmarks against a real CPU.}} \tacorev{\X does not require any changes to support an L2/LLC and optimization modules for an L2/LLC can be flexibly {implemented} similar to the L1. The cost of an MMC miss may be further alleviated with an L2/LLC that reduces access to memory.}

\begin{table}[h]
    \centering
    
    \caption{Parameters of {the} evaluated {real FPGA-based} system.}
    \scriptsize
    \resizebox{\linewidth}{!}{  
    \begin{tabular}{@{} m{40em} @{}}
    \toprule
    \textbf{CPU:} 25~MHz; in-order Rocket core \cite{rocket-chip-gen}; \textbf{TLB} 16 entries DTLB; LRU policy;\\        
    \midrule
	\textbf{L1 Data + Inst. Cache:} 16~KB, 4-way; 4-cycle; 
	64~B line;  LRU policy; MSHR size: 2 \\
    \midrule
    \textbf{MMC:} NMRU Policy; 128 entries; 38bits/entry;  \textbf{Tagging Granularity:} 512B; \\
    \midrule
    \textbf{Private Metadata Table:} 256 entries; 64B/entry; \textbf{DRAM:} 533MHz; $V_{dd}$: 1.5V; \\
    \midrule
    \midrule
    \textbf{Workloads: Ligra~\cite{ligra}:} PageRank (PR), Shortest Path (SSSP), Collaborative Filtering (CF) \\
    Teenage Follower (TF), Triangle Counting (TC), Breadth-First Search (BFS), 
    Radius Estimation (Radii)\\
    Connected Components (CC);  \textbf{Polybench}~\cite{polybench}; \textbf{$\mu$Benchmarks} \\
    \midrule
    \end{tabular}
    }
    \label{table:rocket_params}
\end{table}

\section{Use Case 1: HW-SW Cooperative Prefetching} 
\label{sec:prefetching}

Hardware-software cooperative prefetching techniques have been widely proposed to handle challenging access patterns such as in graph processing~\cite{droplet19,minnow,imp15,ets18,PrefEdge2014,AinsworthGPU16,AinsworthASPLOS18, prodigy-hpca21,tesseract}, pointer-chasing~\cite{DepBasedPref98,CompDir03,CoopPref98,GuidedRegionPref2003,ebrahimi2009techniques,roth1999Effective}, linear algebra computation~\cite{Sunder94} and other applications ~\cite{PrefReinforce2015,DataPrefGraphPrec01,PrefCache2004,NowatzkiISCA19}. In this section, we demonstrate how \X can be flexibly used to implement and evaluate such prefetching techniques. We design a new prefetcher for graph applications that leverages knowledge of the
semantics of graph data structures using \X. Graph applications typically involve irregular pointer-chasing-based
memory access patterns. The data-dependent non-sequential accesses in these
work{loads} are challenging for spatial~\cite{stride,baer2,stride_vector,jouppi_prefetch,ampm,fdp,footprint,sms,sms_mod,spp,vldp,sandbox,bop,dol,dspatch,mlop,ppf,ipcp}, temporal~\cite{markov,stems,somogyi_stems,wenisch2010making,domino,isb,misb,triage,wenisch2005temporal,chilimbi2002dynamic,chou2007low,ferdman2007last,hu2003tcp,bekerman1999correlated,cooksey2002stateless,karlsson2000prefetching} and learning-based hardware prefetchers~\cite{pythiaMicro2021,peled2018neural,peled_rl,hashemi2018learning,shineural,shi2019learning,shineural_asplos,zeng2017long} that rely either on (i) program context information (e.g. program counter, cache line address)  or (ii) memorizing long sequences of cache line addressses to generate accurate prefetch requests.


\emph{To implement the hardware support for our prefetcher, we only needed to add 87 lines of Chisel code to the baseline \X codebase, all within the provided module for new optimization components.} 

\begin{figure}[h]
     \centering
     \includegraphics[width=\linewidth]{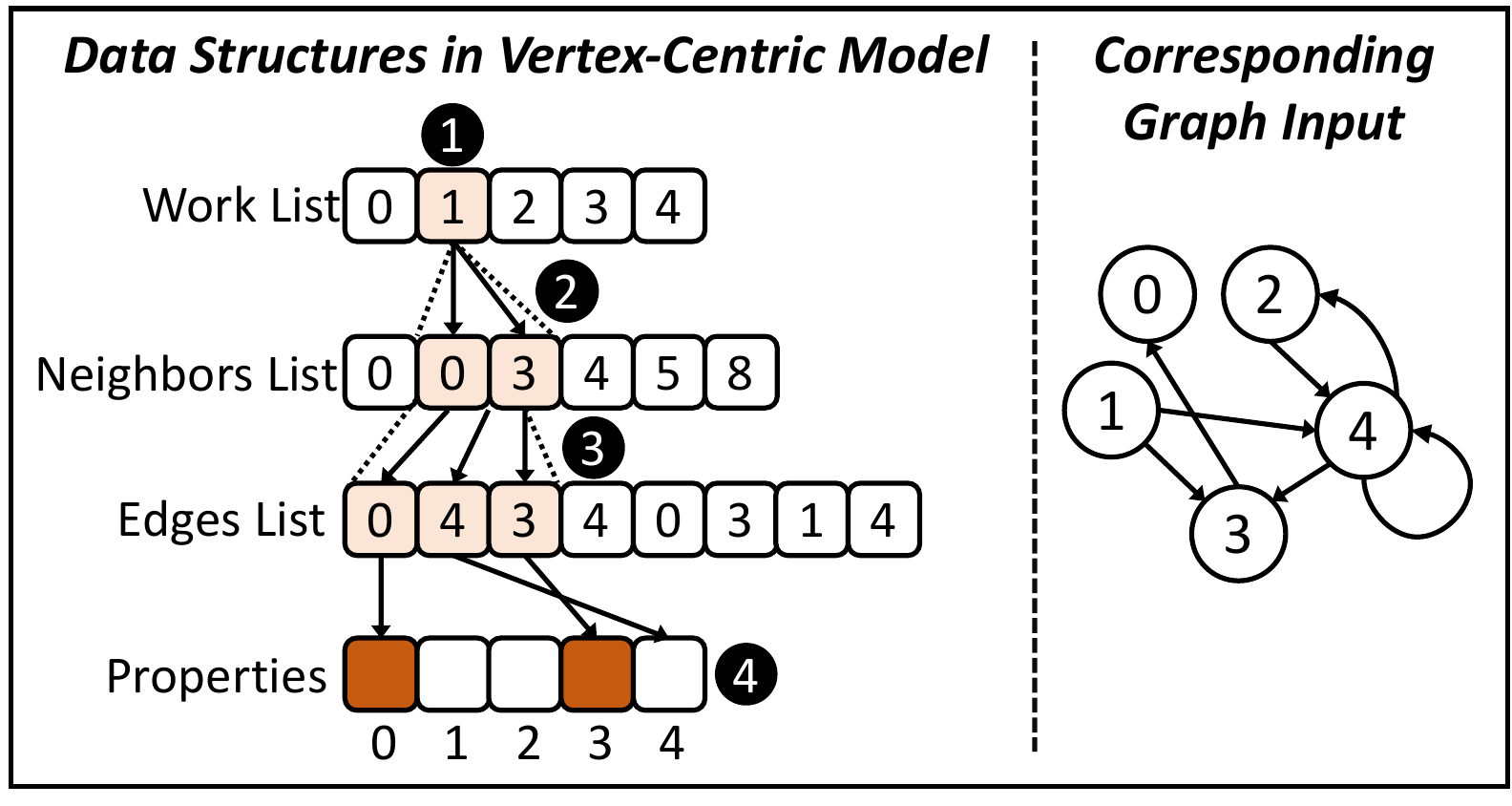}
     
    \caption{Data-dependent accesses in {vertex-centric graph processing model}.}
    
    \label{fig:prefetcher-use-case}
\end{figure}

\mysubsection{Hardware-Software Cooperative Prefetching for Graph Analytics with
\X}
Vertex-centric graph analytics typically involves first traversing a \emph{work
list} containing vertices to be visited (\ding{182} in Fig.~\ref{fig:prefetcher-use-case}). For each
vertex, the {application accesses the} \emph{vertex list} \ding{183} to retrieve the
neighboring vertex IDs from the \emph{edge list} \ding{184}. To perform computation on the
graph, the application then operates on the {properties} of these neighboring
vertices (retrieved from the property list~\ding{185}). Graph
processing thus involves a series of memory accesses that depend on the contents of the
work, vertex and edge lists.

In this use case, we design a prefetcher that can interpret the contents of each
of the above data structures and appropriately compute the next data-dependent
memory address to prefetch. \revision{To capture the required application information
for each data structure, we use \X's \Create interface to communicate the following metadata{: }\One base address of the data structure that is indexed
using the current data structure's contents (64 bits);
\Two base address of the current data structure (64 bits);
\Three data type (32 bits) and size (32 bits) to determine the index of the next access; and \Four the 
prefetching stride (6 bits).} \Map then associates
the address range of each data structure with the appropriate tag. 

{Listing ~\ref{list:metasyspref} shows a detailed end-to-end example of how metadata is created in the application (BFS), how metadata tags are associated with the data structures of BFS{,} and how the prefetcher operates. Lines 3-12 {(incorporated into the code of the BFS application)} use the \X software libraries to create metadata (with \Create) and associate it with the corresponding data structures (using \Map). \Create saves the metadata in the PMT and \Map updates the MMT. Lines 16-34 {(incorporated into the {hardware} optimization client responsible for prefetching)} describe the algorithm behind the MetaSys-based prefetcher. The prefetcher is implemented with an optimization {client (ClientID} = 0). The prefetcher essentially: \One snoops every memory request from
the core and retrieves the associated tag ID using \X; \Two queries the PMT to retrieve the communicated metadata (listed above); and \Three {uses the metadata to identify dependencies between the data structures of the application}.}

{We describe a detailed walkthrough of how the prefetcher operates during the execution of {the} BFS application using Fig.~\ref{fig:prefetcher-use-case} and Listing~\ref{list:metasyspref}.} In Fig.~\ref{fig:prefetcher-use-case}, when the prefetcher {snoops} a memory request {that targets the} work list at index 0, it looks
ahead (depending on the prefetching stride) to retrieve the contents of the
work list at index 1. At this point, it also prefetches the contents of the
vertex, edge, and property lists based on the computed index at each level.
In graph applications where the work list is ordered, the prefetcher is
configured to simply stream through the contents of the vertex and edge lists to
prefetch the data dependent memory locations in the property list. The $snoop\_mem\_request(address)$ (Line 16) function is executed for each request sent by the core to the memory hierarchy. For every memory request, the prefetcher accesses the MMC using the address to receive the tag ID (using MetaSys's lookup functionality). Next, it indexes the PMT using the tag ID to retrieve the metadata associated with the memory request. Using the metadata, the prefetcher determines if the request comes from one of the data structures of the application (Line 24). In this case, the prefetcher first prefetches ahead (Line 26) according to the stride and waits until it receives the value of the prefetched request (Line 28). Using the value, it calculates the address of the data-dependent data structure (e.g., value of WorkList used as an index for VertexList) and looks up the metadata for the newly-formed address. The same procedure happens until no {further} data-dependency is found (Line 22).

The prefetcher can be flexibly {configured (by associating metadata to data structures, Lines 3-12 in Listing~\ref{list:metasyspref}) by the user} based on the specific properties
associated with any data structure, algorithm, {and} the desired
aggressiveness of prefetcher. 

\begin{lstlisting}[caption={\tacorev{Metasys-based Graph Prefetcher. {Available online~\cite{metasysGithub}}.}},captionpos=b,label={list:metasyspref}, float]
/* Additional Code in BFS */
// Create metadata (arguments: ClientID=0, tag ID, metadata)
metadata_create(0, 1, WorkList, { sizeof(Worklist), VertexList, Stride });
metadata_create(0, 2,VertexList, { sizeof(VertexList), EdgeList, Stride});
metadata_create(0, 3,EdgeList, { sizeof(Worklist), VertexList, Stride});
metadata_create(0, 4,Property, {sizeof(Property), NULL, Stride});

// Map tag 1 to Worklist
metadata_map((void*) (WorkList), mapSize, 1);
metadata_map((void*) (VertexList), mapSize, 2); 
metadata_map((void*) (EdgeList), mapSize, 3); 
metadata_map((void*) (Property), mapSize, 4); 

/* Hardware Prefetcher Functionality */
// snoop every memory request
void snoop_mem_request ( address ): 
  // Access MetaSys using the address
  (Valid,Base,Bounds,
   PointerToNextDS,Stride) = metadata_lookup (address); 
    
  // While the data structure traversal is not complete
  while (Valid && PointerToNext != NULL) 
    // If the memory request comes from a tracked data structure 
    if (Base < address && address > Bounds)
      // Initiate a stride prefetch request
      initiate_prefetch(address+stride); 
      // Wait for the prefetch request to return data
      Value = wait_for_value(address+stride);
      // Discover the address of the next data structure (DS)
      address =  &PointerToNextDS[value]; 
      // Look up metadata for the next data structure (DS)
      nextAddress = PointerToNextDS[value];
      (Valid,Base,Bounds,
       PointerToNextDS,Stride) = metadata_lookup (nextAddress); 
\end{lstlisting}

\begin{figure}[h]
    \centering
    \includegraphics[width=\linewidth]{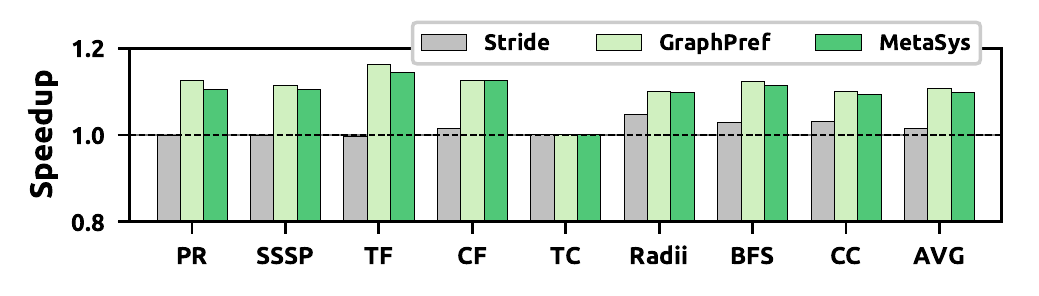}
    
    \caption{Speedup with the \X prefetcher.}
    
    \label{fig:prefetcher-use-case-result}
    
\end{figure}

\mysubsection{Evaluation and Methodology}
We evaluate the \X-based prefetcher using 8 graph {analytics} workloads from the Ligra
framework~\cite{ligra} using the Rocket {Chip} prototype of \X with the system
parameters listed in Table \ref{table:rocket_params}. \revision{We evaluate three configurations:
\One the baseline
system with a hardware stride prefetcher~\cite{stride}; \Two \emph{GraphPref}, a customized hardware prefetcher that implements the same idea described above without the generalized MetaSys support {(similar to prior work~\cite{AinsworthGPU16,prodigy-hpca21})}; and \Three the \X-based
graph prefetcher.} \tacorev{In the case of \emph{GraphPref}, all the required metadata (e.g., base and bound addresses, stride) are directly provided to the prefetcher using specialized instructions. Thus, \emph{GraphPref} is able to access metadata at low latency and does not access the memory hierarchy. The prefetcher works in the same way as the MetaSys-based prefetcher, however, in the case of MetaSys, the general CREATE/MAP instructions are used to communicate information and the metadata lookups {access} the MMC (which may lead to additional memory accesses {when there is} an MMC miss).} \revision{Fig.~\ref{fig:prefetcher-use-case-result} depicts the corresponding speedups, normalized
to the baseline. We observe that the \X graph prefetcher improves
performance by 11.2\% on average (up to 14.3\%) over the baseline by accurately
prefetching data-dependent memory accesses. It also
significantly outperforms the stride prefetcher which is unable to capture the irregular access patterns in graph workloads.
Compared to \emph{GraphPref}, the \X-based prefetcher performs almost as well: within 0.2\% on average ({within} 0.8\% for \texttt{BFS}). \tacorev{The additional overheads of \X come from the MMC misses and the larger number of instructions used.}} \tacorev{In terms of area, \X requires 17KB of SRAM (1KB for the MMC and 16KB for the Private Metadata Table) compared to the custom hardware prefetcher which requires 8KB of SRAM for the metadata. The custom prefetcher require{s} 2 additional instructions and additional logic to perform metadata lookups and create/update metadata. We found the area complexity to be slightly less for the custom solution as the SRAM requirements are lower ($\sim$0.01\% for custom hardware versus $\sim$0.02\% for MetaSys{, compared to a 22nm Intel CPU Core~\cite{intel-22nm-measurement}}). {However, \X{}'s overhead can be amortized over multiple use cases, whereas a custom solution is specific to {a single} use case.}  
}

We conclude that \X can be used to flexibly implement and evaluate 
hardware-software cooperative techniques for prefetching by leveraging \X's
metadata support and interfaces, incurring only small overheads from \X's general metadata management. 

\section{Use Case 2: Memory Safety and Protection}
\label{sec:safety}
We describe two hardware-software cooperative mechanisms for memory safety {and} protection that can be directly implemented with \X. \emph{To implement both use cases, we only add 103 lines of Chisel code to the baseline \X code, all within the new optimization {client that is used in both use cases}.} 

\subsection{Hardware Bounds Checking}
\label{sec:boundschecking}
Unmanaged languages such as C/C++ provide great flexibility in memory management
but an important challenge with these languages is \emph{memory safety}. The
pointer casting and pointer arithmetic supported by these languages allow buffer
overflows and potentially hazardous writes to arbitrary memory locations. Prior
work has demonstrated a range of software approaches~\cite{efficient-backwards-transform, protecting-c-programs, ccured, efficient-detection-pointers,deputy,backwards-compatible-bounds-checking, cyclone, shadow-processing,softbound,purify, practical-overflow-detector,nescheck,mpx-explained,checked-c,mondrian-witchel-asplos02,ext-mondrian} to increase memory safety
in the form of static or dynamic checks, such as CCured~\cite{ccured}, Cyclone~\cite{cyclone}, and Softbound~\cite{softbound}. These approaches are known to incur significant
runtime overheads in performing numerous checks in software~\cite{SoK}.
Hardware-based approaches offer a promising opportunity to alleviate these
overheads. Prior work~\cite{hardbound,lowfat-pointers,watchdog-lite,watchdog,cheri,harmoni,memtracker,shaktiT}, including HardBound~\cite{hardbound}, ShaktiT~\cite{shaktiT}, and Cheri~\cite{cheri} investigate enabling
hardware-software cooperative bounds checking. These approaches require
architectures that are entirely specialized for bounds checking~\cite{hardbound,lowfat-pointers,shaktiT} or
more heavyweight metadata management systems tailored for memory security and
protection~\cite{cheri,memtracker,watchdog,watchdog-lite,harmoni}. In this section, we demonstrate how \X can be used to implement hardware-based bounds checking {at low overhead using} a lightweight and general metadata system. 

\subsubsection{Implementing bounds checking with \X.} 
\tacorev{We use Listing~\ref{list:bounds-checking-code}, where two arrays A and B are traversed with a stride of one element, to illustrate the mechanism.
To implement bounds checking with \X, we use the \texttt{MAP} operator to tag each data structure to be protected with a unique ID (lines {3-5}). For dynamically allocated nodes (which may not be contiguously located), each node is tagged with the \emph{same} ID as other nodes in the same data structure.
Every memory access in the program then needs to be verified in hardware to be going to the correct data structure. To do this, we add the \Create operator before every load or store to a protected data structure (lines {11, 14, and 18}). The \Create operator in this case communicates the tag ID of the desired data structure as metadata and the {Client}ID of the bounds checking hardware. In hardware, we simply check whether there is a match between \Create's tag ID and the ID of the load/store address that follows the \Create instruction. To perform this check, the bounds check optimization client ({ClientID}=0), performs a lookup to the MMC to retrieve the tag ID associated with the load/store address. This ID is compared with the value stored in the PMT by the previous \Create instruction. If there is a mismatch, this indicates a buffer overflow or an access to data that is not part of the intended data structure as the load is accessing data that was not mapped to the same tag ID and using its interface to the OS, \X terminates the program.}


\begin{lstlisting}[caption={{MetaSys-based bounds checking example{. Full source code is available online~\cite{metasysGithub}.}}},captionpos=b,label={list:bounds-checking-code}, float]
/* Example bounds checking software */
// Map TagIDs to three arrays
metadata_map((void*) (array1), mapSize, 1); 
metadata_map((void*) (array2), mapSize, 2); 
metadata_map((void*) (array3), mapSize, 3); 

// Access every element of each array with a stride of one element
for(i = 0 ; i < array_size ; i += 1)
{
    // Create metadata with TagID=1 and Metadata=1 to ClientID=0
    metadata_create(0,1,1);
    int elem1 = array1[i];
    // Create metadata with TagID=2 and Metadata=2 to ClientID=0
    metadata_create(0,2,2); 
    int elem2 = array2[i];
    int result = elem1 + elem2;
    // Create metadata with TagID=3 and Metadata=3 to ClientID=0
    metadata_create(0,3,3); 
    array3[i] = result;
 }

/* Hardware Bounds Checker Functionality */ 
HardwareBoundsChecker(CreateTagID, Address):
    // Software communicates TagID using CREATE
    TagIDRegister <= CreateTagID 
    // Bounds checker client performs a metadata lookup 
    // to find the TagID of address
    MetadataTagID <= PerformMetadataLookup(Address) 
    // Interrupt rocket core if TagIDs do not match 
    // (i.e., access is out of bounds)
    if MetadataTagID != TagIDRegister: 
        InterruptRocketCore()
    
 
\end{lstlisting}

\subsubsection{Methodology and Evaluation}
We evaluate \X-based bounds checking on our prototype with the parameters listed in Table~\ref{table:rocket_params} (tagging granularity
is set to 64B). We use the Olden~\cite{olden} benchmarks (commonly used for bounds checking and stack protection research~\cite{hardbound,cheri,softbound,memsafe,checked-c} due to its focus on pointer-based
data structures). \tacorev{We only compare against a prior software solution~\cite{efficient-backwards-transform} as custom hardware solutions for bounds checking require intrusive changes to the microarchitecture, ISA, and application, which are difficult to reasonably implement on a full-system simulator or an FPGA. For example, {implementing} Hardbound~\cite{hardbound}, the closest custom hardware solution for bounds checking, requires compiler support, extending every register and word of memory with “sidecar” shadow registers for base and bound addresses, compression/decompression engines to compress these base/bound addresses, tag caches, and bounds checking logic.}

\begin{figure}[!h]
    \centering
    \includegraphics[width=\linewidth]{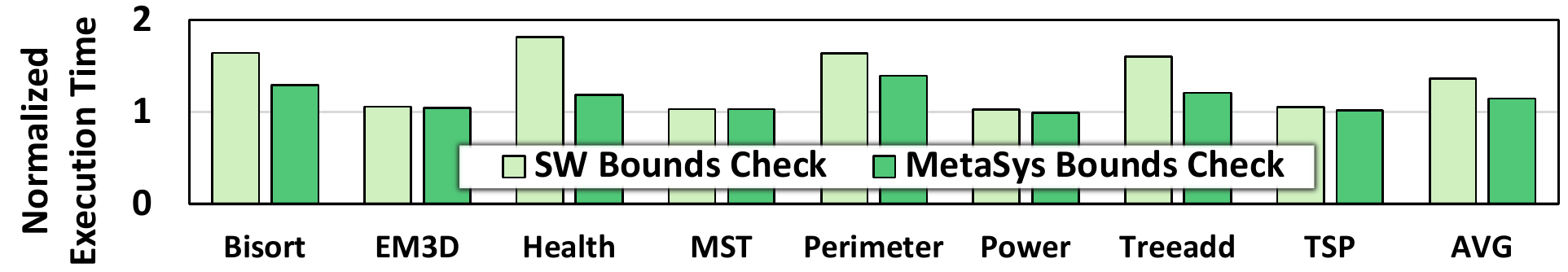}
    \caption{Performance overheads for bounds checking}
    \label{fig:bounds-checking-results}
\end{figure}

We evaluate {three} designs: \One the \emph{Baseline} system without \X; \Two software bounds checking, based on prior work~\cite{efficient-backwards-transform};
and \Three \X-based bounds checking. Fig.~\ref{fig:bounds-checking-results} depicts the execution time normalized to the Baseline. 
We observe that the software bounds checking design incurs
a high {average performance} overhead of 36\% (up to 82\%). This overhead comes from executing more instructions to check bounds (64\% on average). In contrast, \X-based bounds
checking incurs {only an average performance overhead of} 14\% (up to 40\%). \X requires {only} a 32\%
increase in the number of executed instructions. Workloads such as
\texttt{em3d}, \texttt{power}, and \texttt{mst}, are highly compute intensive
and hence do not incur significant overheads with either bounds checking
technique.

We conclude that MetaSys provides a lightweight substrate to
implement and evaluate hardware-software cooperative bounds checking. \X can be
flexibly extended to implement more sophisticated
memory protection techniques.

\subsection{Return Address Protection}
\label{sec:rap}

The program's call stack is a known source of many security vulnerabilities in
low-level, memory-unsafe languages such as C/C++. For example, the control flow
in the program can be hijacked by overwriting the \emph{return addresses} saved
in the stack{~\cite{canaries,smashingforfun}}. Existing defenses such as ExecShield~\cite{execshield} and 
stack canaries~\cite{canaries} do
not protect against sophisticated attack techniques~\cite{rop-wo-return,jump-oriented-programming,return-into-libc-expressiveness}. Stack canary
protection~{\cite{canaries}}
is a software check that involves writing an additional randomly-generated value
in the stack and a duplicate is saved separately in memory. {The stack canary checks the randomly-generated value in the stack against its duplicate to detect stack overwriting before a function returns to the return address saved in the stack.} Protecting
return addresses with more powerful software checks~\cite{shadow-stacks,control-flow-integrity,CCFI,OCFI,cpi} incurs significant runtime
overheads and are hence difficult to use in practice~\cite{shadow-stacks-bad,SoK}. Prior
work has proposed a range of hardware techniques~\cite{raguard,pac-it-up,cheri,protecting-stack,cet,pump} to enable return
address protection more efficiently. These approaches {either} require dedicated
hardware support for stack protection (e.g., RAGuard~\cite{raguard}, PAC-it-up~\cite{pac-it-up}, CET~\cite{cet}) or more heavy-weight metadata systems for memory protection (e.g., SDMP~\cite{protecting-stack}, Cheri~\cite{cheri}, PUMP~\cite{pump}). In this section, we
implement and evaluate return address protection with \X's lightweight metadata
support and cross-layer interfaces.

\vspace{-1mm}
\subsubsection{Return address protection with \X}
\tacorev{To enable return address protection with \X, we first tag each return address
using \Map as id="1". This is done automatically with compiler support and no programmer intervention is required. No \Create instruction is used. In hardware, we add support to simply disallow writes
to any address tagged with id="1". In order to do this, we implement a simple hardware optimization {client} which is triggered by store instructions. For each store instruction, the {client} performs a lookup to the MMC to determine the tag ID associated with the address. If the tag="1", this indicates that the location is a return address and the store is not allowed to complete. Any store to a tagged memory address causes the hardware to
invoke the OS to terminate the program.} The application can then unmap the
return address when it is retrieved again from the stack. This ensures that once a
memory location within the stack has a return address saved, it cannot be
overwritten via attacks that hijack control flow such as buffer overflow attacks.

\vspace{-1mm}
\subsubsection{Evaluation and Methodology.}
We evaluate \X-based return address protection using our FPGA prototype with
system parameters listed in Table~\ref{table:rocket_params} (the tagging granularity
set to 64B). We evaluate 3 designs using the
Olden~\cite{olden} benchmarks: \One the \emph{Baseline} system with no
overheads; \Two canary stack protection~\cite{canaries} in the GCC RISC-V
compiler; and \Three \X-based return address protection.  

\begin{figure}[!h]
    \centering
    \includegraphics[width=\linewidth]{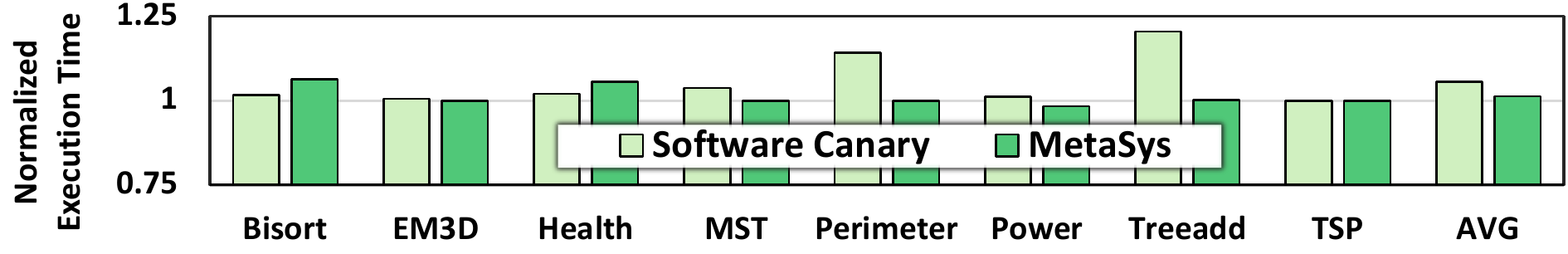}
    \caption{Performance overheads for return address protection}
    \label{fig:return-address-protection-result}
\end{figure}

Fig.~\ref{fig:return-address-protection-result} depicts the execution time normalized to the \emph{Baseline}. We
observe that the canary approach incurs a performance overhead of 5.5\% (up
to 20\%),
while \X incurs a diminished overhead of 1.2\% (up to 6.2\%). The major overheads
for the stack canaries come from executing extra instructions (5.5\% on average) to perform
software checks. \revision{The overheads for \X are low due to the high MMC hit rate which leads to few additional memory
accesses.} In addition to providing
less overhead, \X-based return address protection can also protect
against more sophisticated attacks that exploit write-what-where gadgets~\cite{write-what-where} and, unlike canaries, {are} immune to
information leaks~\cite{breaking-memory-secrecy}. 
Protecting additional memory locations beyond return addresses
(e.g., function pointers) with software approaches
would incur even higher instruction overhead. However, the {observed} \X overhead would
largely remain the same as it already involves checking each
store. {In addition, MetaSys-based return address protection utilizes the Metasys support and interfaces{, whose cost is amortized across many use cases,} without requiring specialized ISA and hardware support.}

We conclude that \X enables easy implementation and evaluation of lightweight memory protection
mechanisms with low
performance overhead.

\vspace{-1mm}
\section{Other Use Cases of \X}
\vspace{-1mm}
\label{sec:use-cases} 
\rev{We briefly discuss {various} other cross-layer techniques that can be implemented with \X (but would be challenging to implement with {prior approaches like} XMem~\cite{xmem}).}

\textbf{Performance optimization techniques.} \X
provides a low-overhead framework and a rich cross-layer interface to implement a diverse set of performance
optimizations including cache management, prefetching, page placement
in memory, approximation, data compression, DRAM cache management, and memory
management in NUMA and NUCA systems~{\cite{swcache-jain-iccad01,compilerpartitioned-ravindran-lctes07,popt-gu-lcpc08,pacman-brock-ismm13,evictme-wang-pact02,generatinghints-beyls-jsysarch05,keepme-sartor-interact05,compilerassisted-yang-lcpc04,cooperativescrubbing-sartor-pact14,
runtimellc-pan-sc15,prefetchtasklifetimes-papaefstathiou-ics13,radar-manivannan-hpca16,modified-tyson-micro95,unlocking-agarwal-hpca15,trafficmanagement-dashti-asplos13,das2013a2c}}. \X can flexibly implement the range of
cross-layer optimizations supported by XMem~\cite{xmem}, {and the Locality Descriptor~\cite{ldesc}}. \X's \emph{dynamic}
interface for metadata communication enables even more powerful optimizations
than XMem including memory optimizations for dynamic data structures such as
graphs. {We already demonstrate one performance optimization in {§}\ref{sec:prefetching}.}

\textbf{Techniques to enforce cross-layer quality of service (QoS).} \X can be used
to implement cross-layer techniques to enforce QoS requirements of applications in shared
environments~\cite{pard-ma-asplos15, labeled-yu, parbs-mutlu-isca08, mutlu2007stall,tcm-kim-micro10,ebrahimi-10,subramanian2013mise,subramanian2015asm}. \X allows communicating an
applications' QoS requirements to
hardware components (e.g., the last-level cache, memory controllers) to enable
optimizations for partitioning and allocating shared
resources such as cache {space} and memory bandwidth.

\textbf{Hardware support for debugging and monitoring.} \X can be used to
implement cross-layer techniques for performance debugging and bug
detection by providing efficient mechanisms to track memory access patterns using {its} memory tagging and metadata lookup support.
This includes efficient detection of memory safety
violations~\cite{memtracker,iwatcher} or concurrency
bugs~\cite{colorsafe,avio,atomtracker,lucia2008atom, conflictexceptions,hard}
such as data races, deadlocks, or atomicity violations. 

\textbf{Security and protection.} \X provides a substrate to implement low-overhead hardware techniques for security/protection: the tagged memory support can be used to implement protection for spatial memory safety~\cite{harmoni,hardware-enforcement,legba,tempest-typhoon}, cache timing side-channels~\cite{timing-attacks} and stack protection~\cite{protecting-stack, pac-it-up}. \highlight{For example, using \X, software can tag memory accesses as security-critical or safe. Based on the metadata received for every access, \X can activate/deactivate (for the specific access) the corresponding side-channel defense technique at runtime (e.g., protect from {or} undo speculation \cite{cleanupspec,invisispec,MI6,barber2019specshield,khasawneh2019safespec}). We {already} demonstrate two security techniques in §\ref{sec:safety}.}

\textbf{Garbage collection.} \X offers an efficient mechanism to track dead
memory regions, unreachable objects, or young objects
in managed languages. \X is hence a natural substrate to implement hardware-software cooperative approaches (such as prior work~\cite{joao2009flexible,maas2018hardware,maas2016grail}) for garbage
collection. \highlight{For example, HAMM~\cite{joao2009flexible}, a hardware-software {cooperative} technique for reference counting, tracks the number of references to any object in hardware. It has many of the same metadata management components as \X. HAMM uses a multi-level metadata cache to manage the large amounts of metadata associated with reference counting for each object. MetaSys was designed with modular interfaces that enable adding more levels to the metadata cache for such use cases.}

\textbf{OS optimizations.} \X can be used to implement OS optimizations that require hardware performance monitoring of memory access patterns, contention, reuse, etc{~\cite{irregularities-park-asplos13, mcp-muralidhara-micro11,stride1,sms,somogyi_stems,datatiering-dulloor-eurosys16}}. The metadata support in \X can {be} used to implement this monitoring and then inform OS optimizations like thread scheduling, I/O scheduling, and page allocation/mapping~\cite{buddy-allocator,buddy-original,cfs,das2013a2c,mcp-muralidhara-micro11}.

\tacorev{\textbf{Cache optimizations.} MetaSys enables various cache optimizations such as cache scrubbing~\cite{evictme-wang-pact02,cooperativescrubbing-sartor-pact14} and cache prioritization~\cite{swcache-jain-iccad01,compilerpartitioned-ravindran-lctes07,popt-gu-lcpc08,pacman-brock-ismm13,evictme-wang-pact02,generatinghints-beyls-jsysarch05,keepme-sartor-interact05,compilerassisted-yang-lcpc04,cooperativescrubbing-sartor-pact14,
runtimellc-pan-sc15,prefetchtasklifetimes-papaefstathiou-ics13,radar-manivannan-hpca16,modified-tyson-micro95}. To implement such optimizations with MetaSys, the \Create operator is used to specify the expected \emph{reuse} of a data object at runtime. For example, objects can be tagged as having no reuse (e.g., once all threads have completed operations on it). Thus, upon encountering a cache miss (the trigger event), the cache controller can look up the expected reuse of different cache lines using MetaSys's lookup mechanism and then evict the dead cache line. A similar mechanism {can be} used to retain cache lines that have high expected cache reuse.}

{\textbf{Compressing sparse data structures.} \X can be used to support techniques that efficiently compress sparse data structures and accelerate sparse workloads~\cite{smash,seshadri2015page}. For example, SMASH~\cite{smash} {is} a hardware-software {cooperative technique} that efficiently compresses sparse matrices using a hierarchy of bitmaps to encode non-zero cache lines and accelerates the discovery of the non-zero elements of the sparse matrix. Instead of using specialized hardware, SMASH could access the hierarchy of bitmaps and identify non-zero elements with \X{}' metadata support.}

{\textbf{Heterogeneous reliability memory optimizations.} \X{}' metadata support can be used by techniques that exploit heterogeneous reliability characteristics of memory devices to improve performance{,} power consumption{, and system cost}~\cite{eden,flikker-liu-asplos11,approximate-sampson-tocs14,characterizing-luo-dsn14}. These techniques typically require support for dynamically looking up the {error tolerance} characteristics of data structures to place {them} in memory to {satisfy a target bit error rate}. \X{}' metadata support is a natural candidate for providing these techniques with a means to {query reliability} characteristics of data structures.}
\section{Limitations of \X}

Our goal of providing a low-overhead and general system largely tailored for cross-layer \emph{performance} optimization leads to {several major} limitations in \X. {These limitations can be mitigated by future work.}

\textbf{Instruction and register tagging {are not supported}.} \X does not currently support tagging of instructions or registers and thus cannot easily support techniques such as taint-tracking~\cite{hdfi, minos, spe} and security mechanisms that require rule-checking at the instruction/register level~\cite{software-defined-metadata,pump}. 

\textbf{{Overheads of f}ine-granularity memory tagging.} While \X supports memory tagging at flexible granularity, the system is optimized for the \emph{larger} granularities typically required for performance optimization ($>=$64B) \revision{or fine granularities for only \emph{some} data (e.g., return addresses). Byte/word granularity tagging for the entire program data may lead to high MMC miss rates and may thus incur higher overheads with \X.} 

\tacorev{\textbf{{Limitations on u}sing {private metadata tables (PMTs)} for runtime profiling.} With MetaSys's existing interfaces, the PMTs cannot be used to collect program information and supply it back to the application. The PMTs can only be {updated by} the \Create operator. {This issue can be mitigated relatively easily {in} future {MetaSys} versions.}}

\section{Characterizing General Metadata Management Systems for Cross-Layer Optimizations}
\label{sec:tradeoffs}
Our goal in this section is to perform a detailed characterization of the overheads of using a \emph{single} common metadata system and interface for multiple cross-layer techniques. Three major challenges and sources of system overhead include:

\noindent
\emph{(1) Handling dynamic metadata:} Communicating metadata at runtime
requires execution of additional instructions in the program. This incurs performance overheads in
the form of CPU processing cycles and data movement to communicate the metadata
to hardware components or to save them in memory. 

\noindent
\emph{(2) Efficient metadata management and lookups:} The communicated
metadata must be saved in memory or specialized caches (the MMC in \X) that overflow to memory. Different components in the system must then be able to efficiently look up the
metadata for performance optimization. Storing and retrieving metadata may
incur expensive memory accesses and consume memory bandwidth.

\noindent
\emph{(3) Scaling to multiple components:} A \emph{general} cross-layer
interface and metadata system must be able to serve multiple client components
implementing different optimizations in the caches, 
the prefetchers, the memory controller, etc. Multiple
components accessing shared metadata support during program execution poses significant
scalability challenges.

{T}he above challenges may impose significant area and performance overheads
in the CPU, making the feasibility of a common metadata system and interface (as opposed to per-use-case specialized interfaces and {optimizations}) for cross-layer techniques questionable. In this section, we set out to experimentally quantify these overheads, identify key bottlenecks, and {discover and provide} insights on how these challenges affect different workloads and how they can be alleviated.

\subsection{Analysis}

\label{sec:analysis}
We perform our characterization using the Polybench~\cite{polybench} and
Ligra~\cite{ligra} benchmark suites along with a set of
microbenchmarks {(available on Github~\cite{metasysGithub})}. Polybench contains building block
kernels frequently used in linear algebra, scientific computation, and
machine learning. Ligra contains widely used graph analytics workloads.
The microbenchmarks are designed to \emph{intensely} stress the \X system and identify \emph{worst-case} overheads.
\begin{itemize}
\item \emph{Stream.} This memory-bandwidth-intensive microbenchmark streams through a large
amount of data, accessing it only once. It hence has high spatial locality and no data reuse.
\item \emph{Linked List Traversal.} The microbenchmark mimics
typical linked list creation, insertion, and traversal and emulates the widely-seen
memory-intensive \emph{pointer-chasing} operation. 
\item \emph{Random Access.} This microbenchmark accesses memory locations within
a large array at random indices and
is designed to {test an extremely rare worst-case scenario:} no pattern in
accesses, no reuse, and no spatial locality. 
\item
\emph{3-dimensional array traversal (3D Array).} This microbenchmark mimics the access pattern and locality seen in
applications with multi-dimensional arrays.
It traverses a 3D array first along the third
dimension, and then along the second and first (data is contiguously placed in the first dimension). The access pattern is highly
regular but {exhibits} no
spatial locality.\end{itemize}

§\ref{sec:methodology} describes the parameters of our baseline system. We summarize our key findings
in §\ref{subsec:findings}. In all evaluations in this section, since we aim to characterize the overheads of the system itself, we do
\emph{not} implement any cross-layer optimization that improves performance. We simply implement \emph{lookups} to the metadata system that an optimization {could potentially} make. Since our goal is to stress the system {and understand the worst-case overheads}, we perform {metadata} lookups for \emph{every
memory access}. In typical use cases, the lookup trigger{s}
would be {much} less frequent {than for every memory access}, e.g., lookups on {only} cache miss{es} for prefetching
or {only} store{s} for return address protection. 

\subsubsection{Performance Overhead Analysis}
The performance
overheads in \X come
from two major sources: \emph{(i)} dynamic instructions (\texttt{MAP} and
\Create) and \emph{(ii)}
metadata \emph{lookups} when a component retrieves the tag ID associated with any memory
address (from the MMT, cached in the MMC) and then the corresponding metadata (in the
PMT).

\begin{figure}[h]
	\centering
	
	\includegraphics[width=\linewidth]{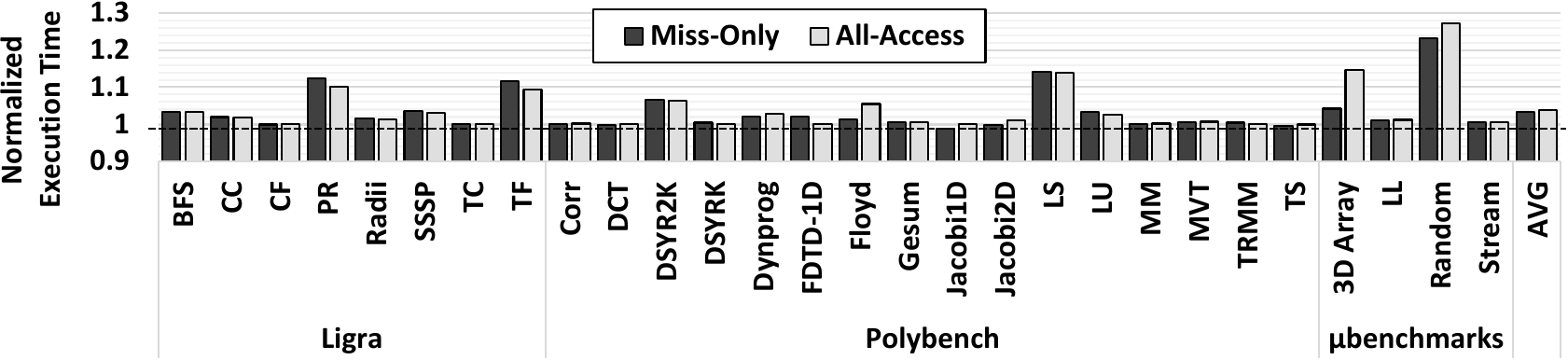}
	
	\caption{Normalized performance overhead {of MetaSys.}}
	\label{fig:overhead-baseline}
	
\end{figure}

\begin{figure}[h]
	\centering
	\includegraphics[width=\linewidth]{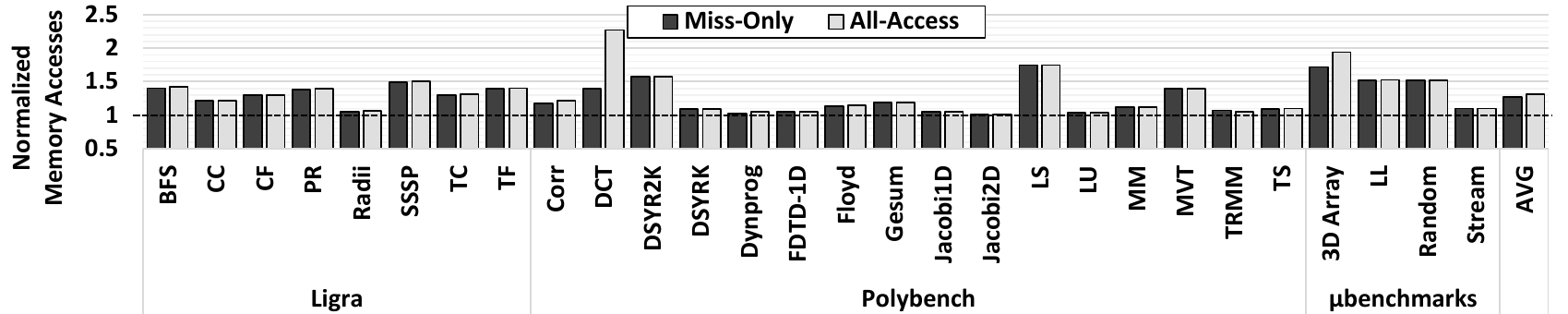}
	 
	\caption{Additional memory accesses introduced by {MetaSys metadata} lookups.}
	\label{fig:memory-accesses}
\end{figure}

\begin{figure}[h]
	\centering
	\includegraphics[width=\linewidth]{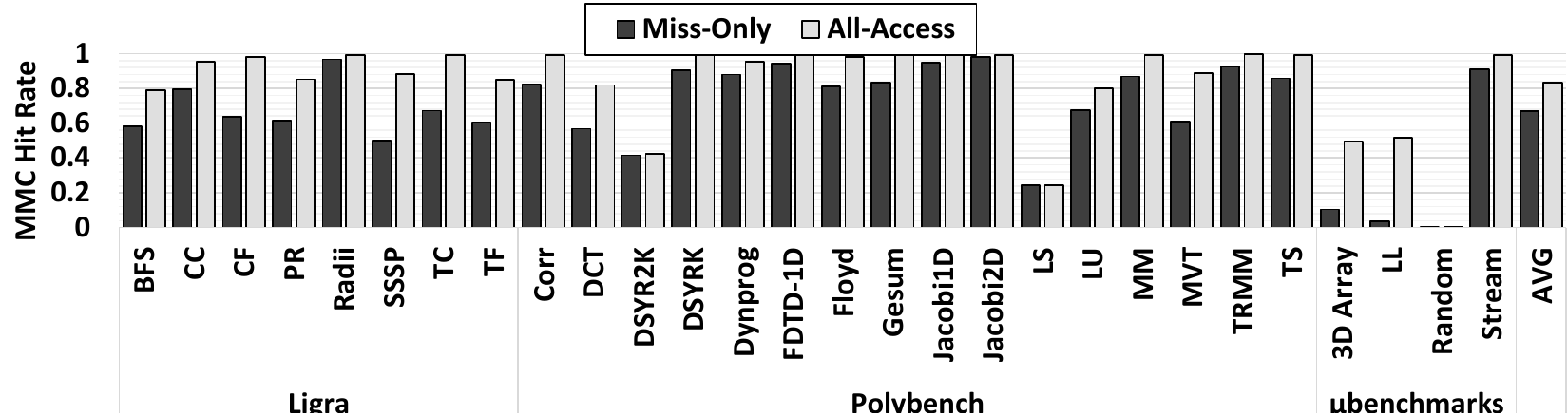}
	
	\caption{Metadata Mapping Cache (MMC) hit rate.}
	\label{fig:alb-hit-rate}
\end{figure}

Fig.~\ref{fig:overhead-baseline} depicts the execution time normalized to the
baseline system (without \X) for two scenarios: \One when
performing metadata lookups for \emph{every} access to the L1 cache
(\emph{All-Accesses}) and \Two
when performing metadata lookups only on every L1 cache miss (\emph{Miss-Only}). These studies
were conducted with our baseline 128-entry MMC with a tagging granularity of 512B (as
in XMem~\cite{xmem}). 
Fig.~\ref{fig:memory-accesses}
plots the number of memory accesses, normalized to baseline (the additional memory accesses come from misses in the MMC). {Fig.~\ref{fig:alb-hit-rate}
plots the corresponding MMC hit rates.
}


We make two major observations from the {three} figure{s}. First, the overall {performance} overheads
from the metadata management system for both designs are low in most workloads with an average {performance}
overhead of 2.7\% (ranging from $\sim$0\% up to 14\%), excluding the
microbenchmarks. The highest overheads observed in the microbenchmarks is 27\%
for \texttt{Random} and represents the worst-case overhead. Workloads with the highest
overheads (\texttt{Random}, \texttt{GS}, \texttt{PR}, \texttt{TF}) are highly memory-intensive and
have low spatial and temporal locality, which leads to
low hit rates in the MMC (e.g., $\sim$0\% in \texttt{Random} and
24\% in \texttt{GS}). This causes a significant increase in accesses to
memory and thus higher performance overheads. 

Second, the \emph{number} of metadata lookups {(not shown in the figure)} {does not have a direct impact} on the overall
performance overhead. \emph{All-Access} 
performs on average 75.2\% more lookups than \emph{Miss-Only}, but incurs an
additional overhead of only 0.05\%. \emph{Miss-Only} has lower MMC hit rates due
to lower locality in lookups than
\emph{All-Access}. Thus, the number of \emph{additional} memory accesses is largely
the same for both designs, {as shown in Fig.~\ref{fig:memory-accesses}}.

\begin{figure}[!h]
    {\centering
    
    \includegraphics[width=\linewidth]{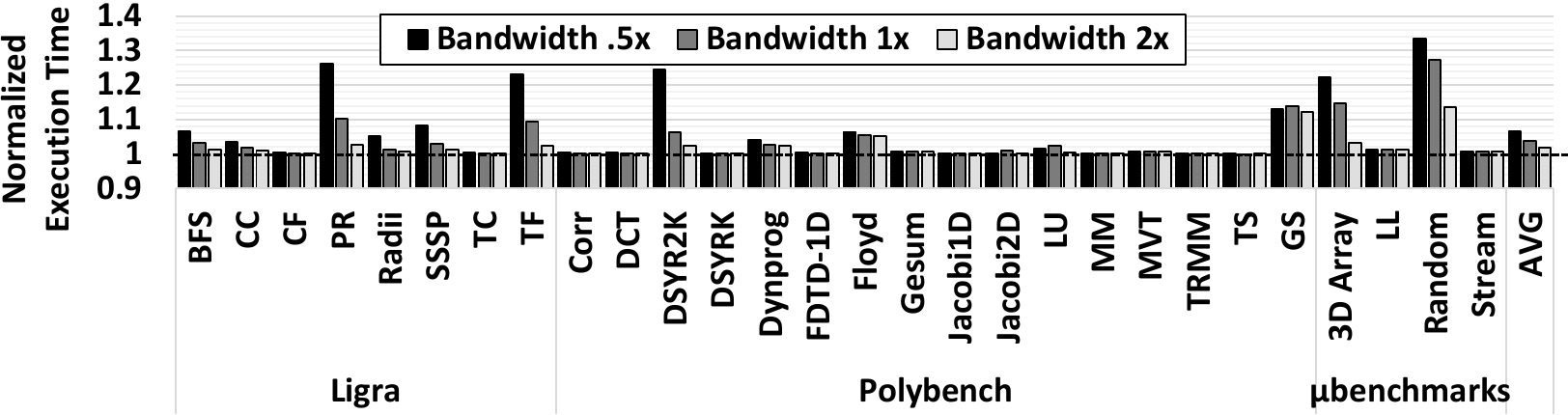}
    
    \caption{{MetaSys performance overhead on systems with varying amounts of memory bandwidth.}}
    \label{fig:bandwidth}
    }
    
\end{figure}

Since the major overheads are from additional memory accesses, we evaluate the
impact of available memory bandwidth. Fig.~\ref{fig:bandwidth}
depicts the performance overhead of \emph{All-Access} {on two different systems} with 0.5$\times$ and
2$\times$ the memory bandwidth {of} the baseline system. We observe that,
except for \texttt{GS}, more memory bandwidth
significantly reduces any lookup overheads: the average overhead is only 0.5\% {on a system with} 2$\times$ the {memory} bandwidth. Conversely, in workloads with higher
MMC miss rates (e.g, \texttt{DSYR2K}), {performance} overheads increase with a reduction
in available memory bandwidth.

We conclude that \One performance overheads are correlated to the
MMC hit rates; \Two metadata lookup hardware can
be frequently queried with no direct observable impact on performance; and \Three overall performance overheads are {small} when the MMC
provides high hit rates. 

\begin{figure*}[!b]
    {\centering
        
    \includegraphics[width=1.0\textwidth]{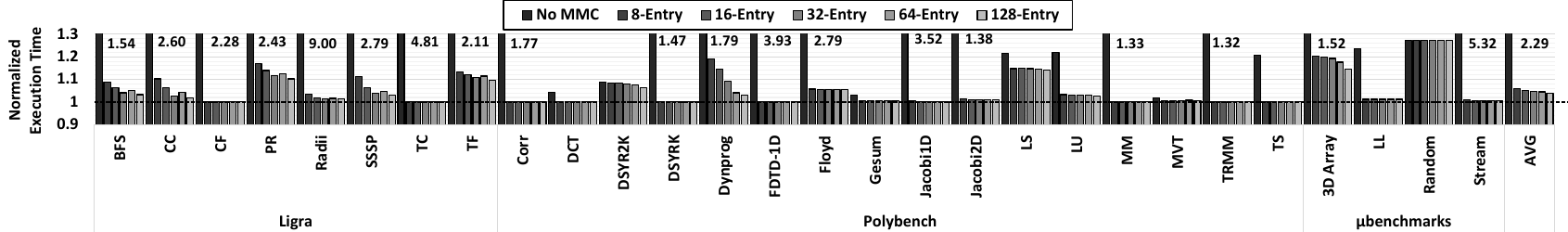}
    
    \caption{Impact of Metadata Mapping Cache (MMC) size on {MetaSys} performance overhead.}
    \label{fig:albsize-exe}
    
}
\end{figure*}
            
\begin{figure*}[!b]

{\centering
\includegraphics[width=1.0\textwidth]{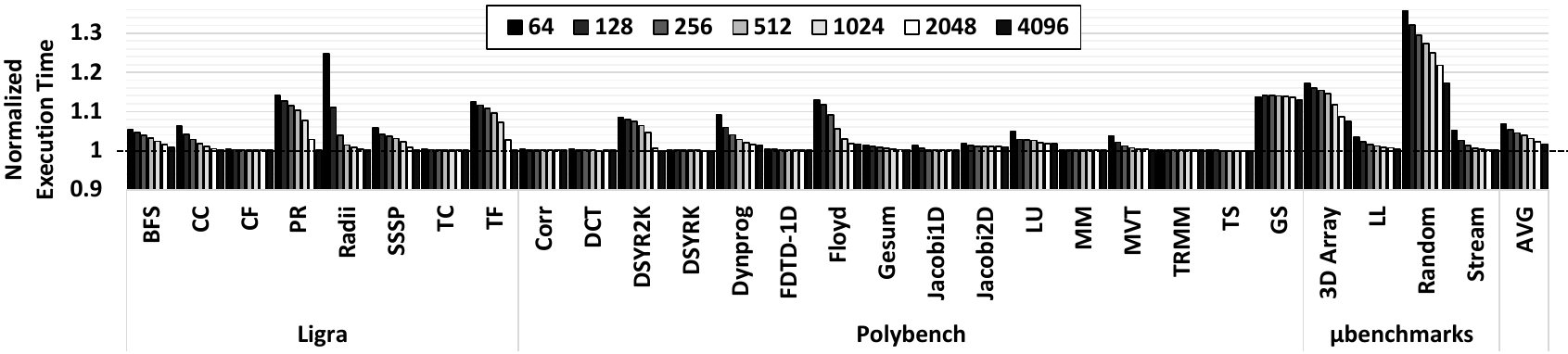}
\caption{Impact of tagging granularity on {MetaSys} performance overhead.}
\label{fig:granularity-exe}
}

\end{figure*}

\subsubsection{Effect of the Metadata Mapping Cache (MMC)}
Fig.~\ref{fig:albsize-exe} {shows} the impact of the size of the MMC on performance overhead. We evaluate 6 sizes for \emph{All-Access} and Fig.~\ref{fig:albsize-exe}
presents the resulting execution time (normalized to the baseline system). We make two observations. First, in most
workloads, 128 entries is sufficient to obtain {small} overheads. This is because
with a 512B tagging granularity, we can hold tag IDs for 64KB of memory in the MMC (compared to 16KB of L1 cache space). Second, workloads with
poor spatial and temporal locality (e.g., \texttt{Random}, \texttt{GS}), are largely
insensitive to the MMC sizes evaluated. {Thus,} overheads {in such workloads} cannot be easily
addressed by increasing {the MMC} size.

\begin{figure*}[t]
{\centering
\includegraphics[width=1.0\linewidth]{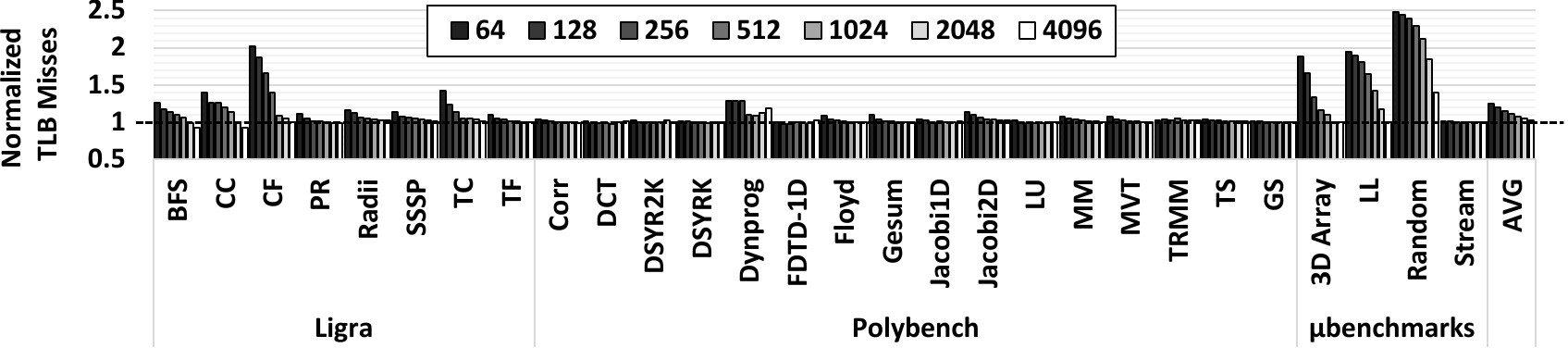}
\caption{Impact of tagging granularity on {additional} TLB misses.}
\label{fig:granularity-tlb}}

\end{figure*}

\subsubsection{Effect of Metadata Granularity}
The \emph{granularity} at which memory is tagged plays a critical role in
determining the reach of the MMC. {Tagging memory at smaller granularities requires more MMC entries to tag the same amount of memory, but it enables more optimizations (e.g., bounds checking).}
Fig.~\ref{fig:granularity-exe} presents execution time for different
granularities of tagging, normalized to the baseline system without \X. 
For most workloads, even the
smallest granularity we evaluated (64B) has a {small} impact on performance.  
{Large} granularities {reduce} overheads for all but \texttt{Random}
and \texttt{GS} by significantly increasing the MMC hit rate. 
A secondary effect in irregular
workloads, such as \texttt{PR} and \texttt{SSSP}, is that {small} granularities increase the number of \emph{TLB} misses (by 11\% and 13\% respectively), {as} depicted in Fig.~\ref{fig:granularity-tlb}. The additional MMC misses cause accesses to the MMT in memory which requires address translation.

\begin{figure}[h]
        {\centering
        \includegraphics[width=1.0\linewidth]{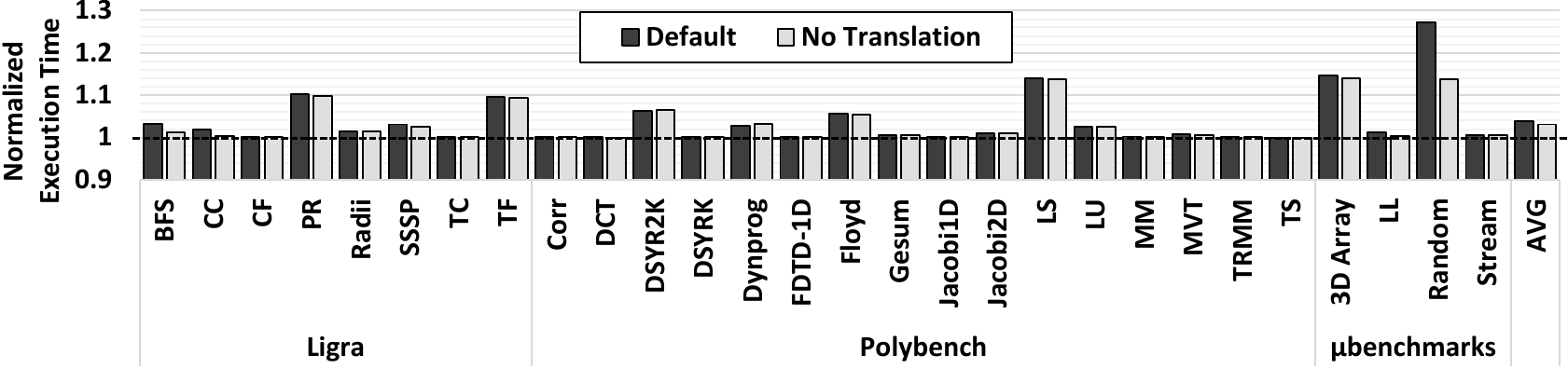}
        \caption{Performance overhead with no address translation {overhead} for metadata.}
        
        \label{fig:physical-table-exe}}
        \end{figure}

To evaluate the effect of the TLB, we implement a {\X{}} design {that} does \emph{not}
require address translation to access the MMT (i.e., {MMT entries are accessed} directly using {physical addresses}). Fig.~\ref{fig:physical-table-exe} presents the resulting normalized execution time
without address translation. We observe a decrease in overhead with this design in the irregular workloads: \texttt{BFS}, \texttt{CC}, \texttt{Random}, and \texttt{LL} (by 1.9\%, 1.8\%, 14\%, and 1\% respectively).

\subsubsection{Effect of Contention {for Metadata}}
To evaluate the \emph{scalability} of {\X{}} with multiple {clients} accessing the same metadata support, we evaluate the overheads of two {clients} performing frequent metadata lookups: one {client} {on} every \emph{memory access} (with the corresponding memory address) and another {on} every \emph{TLB miss} (with the page table entry address). Since each design performs lookups with \emph{different} memory addresses, they do not share entries in the MMC and this creates a {difficult} scenario for the shared MMC.

\begin{figure}[!ht]
        {\centering
       
        \includegraphics[width=1.0\linewidth]{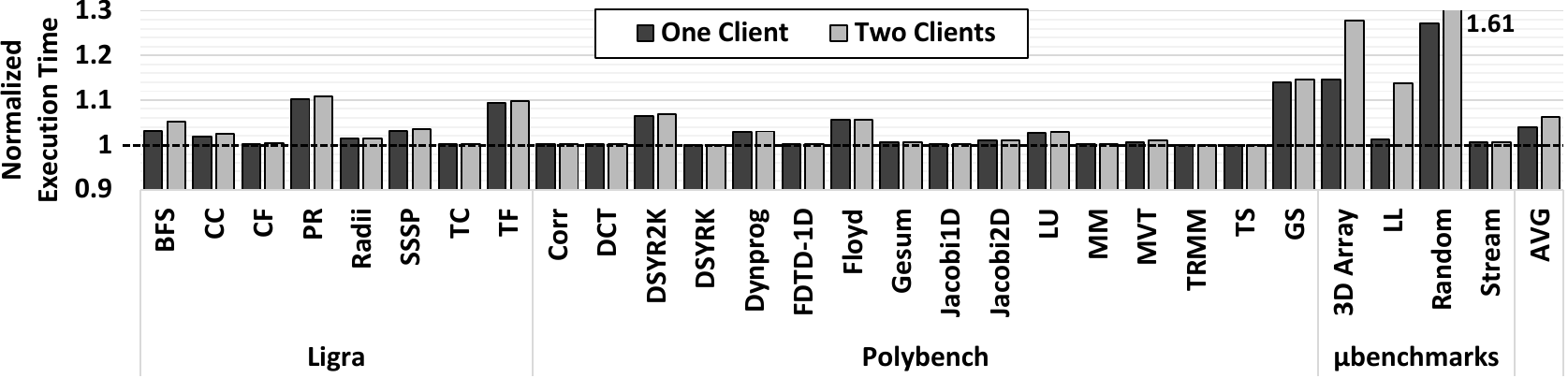}

        \caption{Performance overhead with multiple clients.}
        
        \label{fig:contention-exe}
        }          
        \end{figure}

Fig.~\ref{fig:contention-exe} depicts the resulting execution time for {two} designs, normalized to the baseline system: {\One One Client performs metadata lookups {on} every memory access and \Two Two Clients performs metadata lookups {on} every memory access {as well as on every TLB miss}.} We observe that for all workloads except the microbenchmarks, increasing the number of client{s} {leads to a small additional} performance overhead (on average 0.3\%). This is because the MMC can sufficiently capture the tag ID working set for both {clients}. The microbenchmarks designed to stress the system {experience a} significant {additional} performance {overhead} (up to {3{4}\%} {for Random}) as a result of more misses in the MMC {due to more clients}. 
    
        \begin{figure}[ht]
        {\centering
        
        \includegraphics[width=1.0\linewidth]{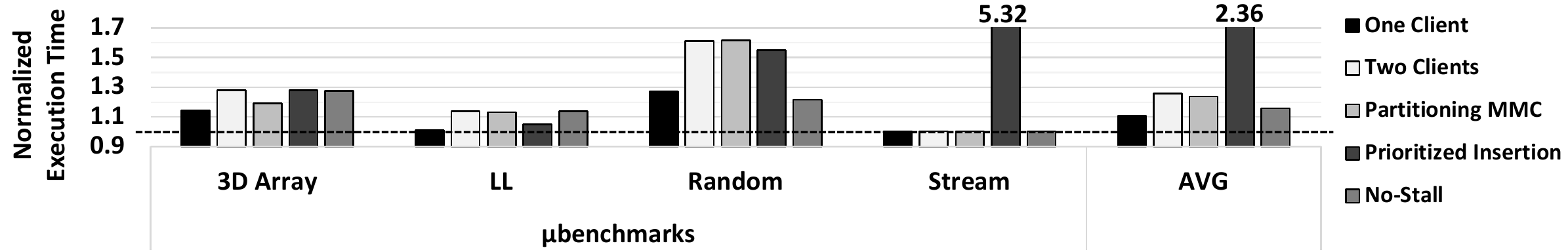}
        \caption{Alleviating MMC contention in microbenchmarks.}
        
        \label{fig:contention-opt}}
        
        \end{figure}
   
To investigate mechanisms to alleviate the {MMC} contention overheads seen in the microbenchmarks, we evaluate three designs in Fig.~\ref{fig:contention-opt}: \One \emph{Partitioning} the MMC equally between the two clients; \Two \emph{Prioritized Insertion}, where we insert mappings for the client with better locality at a higher priority in the MMC ({such that such mappings} are evicted last); and \Three \emph{No stall}, where we do not stall the core on an MMC miss ({instead,} the optimization performed by the client {is} delayed). We observe that \emph{Partitioning} reduces the overhead for \texttt{3D Array} and \texttt{LL} by 9\% and 4\% by avoiding cache thrashing. \emph{Prioritized Insertion} helps reduce the overheads in \texttt{LL} (by 8.5\%) and \texttt{Random} (by 6\%), where one client has more locality than the other in lookups. \emph{No Stall} significantly reduces the overhead in \texttt{Random} (by 40\%) by mitigating the latency overhead of additional memory accesses.

\begin{figure}[h]
    {
    \centering
    \includegraphics[width=1.0\linewidth]{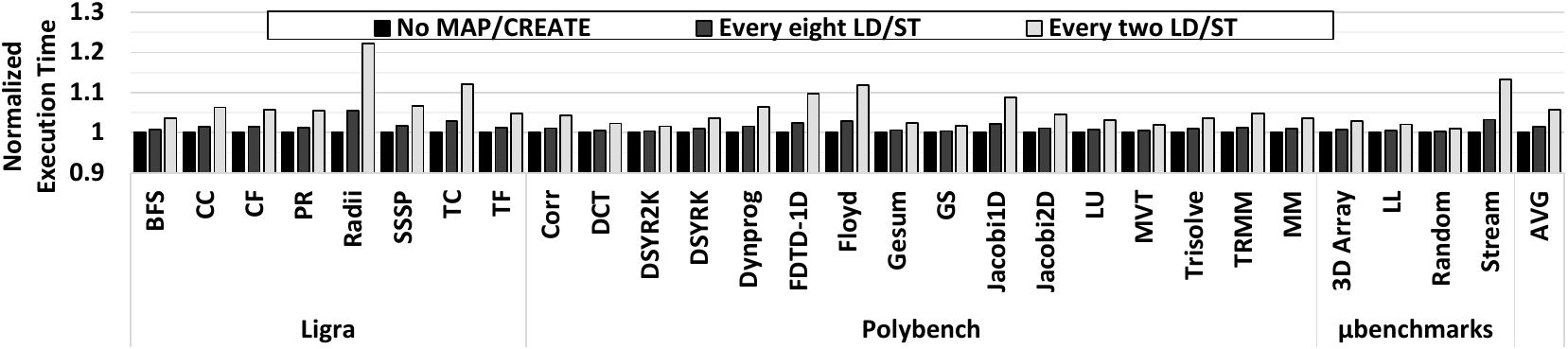}
    \caption{{Performance overhead of MAP/CREATE instructions.}}
    \label{fig:inst-overhead}
    }
    \end{figure}

\tacorev{\subsubsection{Evaluating instruction overheads} To evaluate the instruction overheads of the dynamic \Map/\Create instructions, we present an analysis in Fig.~\ref{fig:inst-overhead} where we intensively use these instructions: \One for every {eight} memory instruction{s,} we add {one} \Map and {one} \Create instructions and \Two for every two memory instructions, we add one \Map and one \Create instruction. We observe {{an} average} slowdown of 1.4\% and 5.7\%, respectively {over a baseline without \X{}}. This indicates that while {excessive} use of \Map/\Create instructions can lead to slowdowns, the primary overheads are still from the metadata lookups which may lead to additional memory accesses.}

We conclude that {MetaSys'} metadata support is scalable to multiple components with {small} impact on performance overheads (except in microbenchmarks). The overheads seen in microbenchmarks are a result of poor MMC hit rates that can be mitigated via techniques such as partitioning, prioritized insertion, and by not stalling the core on an MMC miss. \revision{Since optimizations are triggered by loads/stores in MetaSys, {MetaSys} can be expected to {gracefully} scale to more than two clients as most clients {are expected to} query the MMC {using} the same addresses, which are aggregated and thus {would} not lead to {additional} lookups.}

\subsection{Hardware Area Overhead}

We synthesized the baseline \X system using the Synopsys DC~\cite{synopsys} at 22nm process technology to estimate the area overhead. \X incurs small area overhead: 0.03$mm^2$ (0.02\% of a 22nm Intel {Ivy Bridge} CPU Core~\cite{intel-22nm-measurement}).

\subsection{Summary of Findings}

\label{subsec:findings}

(1) Despite stressing the metadata support, the overall {performance} overheads of {MetaSys} are very low (2.7\% on average, excluding the microbenchmarks). This indicates that using metadata systems
that are \emph{general} enough to support a range of use cases is a promising
approach {to enabling} cross-layer performance optimizations in {a general-purpose manner in} real-world applications. The higher overheads seen in microbenchmarks indicate that the worst-case overheads are however substantially higher (up to 27\%).

(2) Our studies indicate that {MetaSys'} metadata management is \emph{scalable} {to} support multiple client components that have high {rates of metadata access requirements}. {Performance overhead is dependent on locality of metadata accesses as opposed to number of metadata accesses,} indicating that
the same system can support multiple cross-layer optimizations at the same time. We propose simple techniques to alleviate {metadata} contention {generated by} multiple clients. 

(3) The most critical factor that {impacts} {MetaSys'} performance overhead
{is} the effectiveness of the {Metadata Mapping Cache (MMC)}. Workloads with low locality in metadata lookups incur performance overheads
from additional memory accesses. The \emph{reach} of the MMC is also
affected by the \emph{granularity} at which memory is tagged and hence the MMC
hit rate can be improved with larger granularities. Thus, efficient caching of metadata tags is critical.

(4) {A}ccesses to {metadata mappings} in memory require address
translation and cause TLB misses, leading to high performance overhead {especially in irregular workloads (e.g., Random)}. We {find} that this {overhead} can be
mitigated by {using physical addresses to access metadata mappings} or by {storing address translations required to access metadata mappings in a separate TLB.}

\section{Conclusion}

This work introduces \X, {the first} open-source full-system FPGA-based infrastructure to rapidly implement and evaluate diverse cross-layer optimizations in real hardware. We demonstrate \X's versatility and ease-of-use by implementing and evaluating three new cross-layer techniques. We believe and hope \X can enable new ideas and their rigorous evaluation on real hardware.

Using \X, we present the first detailed experimental characterization to evaluate the efficiency and practicality of a single metadata system for cross-layer performance optimization. We demonstrate that the associated performance and area overheads are small, identify key performance bottlenecks, and propose simple techniques to alleviate them. Our characterization thus indicates that a \emph{general} hardware-software interface with lightweight metadata management support offers a promising approach towards enabling {general-purpose} cross-layer {techniques} in CPUs.

\section*{Acknowledgements}
{We thank the anonymous reviewers of {TACO,} MICRO 2020, ISCA 2020{/2021}, {and} SIGMETRICS 2020 for feedback{. We thank} the SAFARI group members for feedback and the stimulating intellectual environment they provide. We acknowledge the generous gifts provided by our industrial partners: Google, Huawei, Intel, Microsoft, VMware, and {the ETH Future Computing Laboratory}.}

\balance
\begin{normalsize}
 \bibliographystyle{IEEEtranS}

 \bibliography{paper,ldesc}

\begin{thebibliography}{100}
\providecommand{\url}[1]{#1}
\csname url@samestyle\endcsname
\providecommand{\newblock}{\relax}
\providecommand{\bibinfo}[2]{#2}
\providecommand{\BIBentrySTDinterwordspacing}{\spaceskip=0pt\relax}
\providecommand{\BIBentryALTinterwordstretchfactor}{4}
\providecommand{\BIBentryALTinterwordspacing}{\spaceskip=\fontdimen2\font plus
\BIBentryALTinterwordstretchfactor\fontdimen3\font minus
  \fontdimen4\font\relax}
\providecommand{\BIBforeignlanguage}[2]{{%
\expandafter\ifx\csname l@#1\endcsname\relax
\typeout{** WARNING: IEEEtranS.bst: No hyphenation pattern has been}%
\typeout{** loaded for the language `#1'. Using the pattern for}%
\typeout{** the default language instead.}%
\else
\language=\csname l@#1\endcsname
\fi
#2}}
\providecommand{\BIBdecl}{\relax}
\BIBdecl

\bibitem{control-flow-integrity}
M.~Abadi, M.~Budiu, U.~Erlingsson, and J.~Ligatti, ``{Control-flow
  Integrity},'' in \emph{CCS}, 2005.

\bibitem{unlocking-agarwal-hpca15}
N.~Agarwal, D.~Nellans, M.~O'Connor, S.~W. Keckler, and T.~F. Wenisch,
  ``{Unlocking Bandwidth for {GPUs} in {CC-NUMA} Systems},'' in \emph{HPCA},
  2015.

\bibitem{pageplacement-agarwal-asplos15}
N.~Agarwal, D.~W. Nellans, M.~Stephenson, M.~O'Connor, and S.~W. Keckler,
  ``{Page Placement Strategies for {GPU}s within Heterogeneous Memory
  Systems},'' in \emph{ASPLOS}, 2015.

\bibitem{tesseract}
J.~Ahn, S.~Hong, S.~Yoo, O.~Mutlu, and K.~Choi, ``{A Scalable
  Processing-in-Memory Accelerator for Parallel Graph Processing},'' in
  \emph{ISCA}, 2015.

\bibitem{AinsworthGPU16}
S.~Ainsworth and T.~M. Jones, ``{Graph Prefetching Using Data Structure
  Knowledge},'' in \emph{ICS}, 2016.

\bibitem{AinsworthASPLOS18}
S.~Ainsworth and T.~M. Jones, ``An event-triggered programmable prefetcher for
  irregular workloads,'' in \emph{ASPLOS}, 2018.

\bibitem{CompDir03}
H.~Al-Sukhni, I.~Bratt, and D.~A. Connors, ``{Compiler-Directed Content-Aware
  Prefetching for Dynamic Data Structures},'' in \emph{PACT}, 2003.

\bibitem{smashingforfun}
\BIBentryALTinterwordspacing
{Aleph One}, ``{Smashing The Stack For Fun And Profit},'' 1996. [Online].
  Available:
  \url{https://inst.eecs.berkeley.edu/~cs161/fa08/papers/stack_smashing.pdf}
\BIBentrySTDinterwordspacing

\bibitem{DataPrefGraphPrec01}
M.~Annavaram, J.~M. Patel, and E.~S. Davidson, ``{Data Prefetching by
  Dependence Graph Precomputation},'' in \emph{ISCA}, 2001.

\bibitem{rocket-chip-gen}
K.~Asanovi{\'c}, R.~Avizienis, J.~Bachrach, S.~Beamer, D.~Biancolin, C.~Celio,
  H.~Cook, P.~Dabbelt, J.~R. Hauser, A.~M. Izraelevitz, S.~Karandikar,
  B.~Keller, D.~Kim, J.~Koenig, Y.~Lee, E.~Love, M.~Maas, A.~Magyar, H.~Mao,
  M.~Moret{\'o}, A.~Ou, D.~A. Patterson, B.~H. Richards, C.~Schmidt, S.~M.
  Twigg, H.~Vo, and A.~Waterman, ``{The Rocket Chip Generator},'' Technical
  Report: UCB/EECS-2016-17, 2016.

\bibitem{efficient-detection-pointers}
T.~M. Austin, S.~E. Breach, and G.~S. Sohi, ``{Efficient Detection of All
  Pointer and Array Access Errors},'' in \emph{PLDI}, 1994.

\bibitem{zedboard}
\BIBentryALTinterwordspacing
AVNET, ``Zynq®-7000 zedboard,'' 2021. [Online]. Available:
  \url{http://zedboard.org/product/zedboard}
\BIBentrySTDinterwordspacing

\bibitem{chisel}
J.~Bachrach, H.~Vo, B.~Richards, Y.~Lee, A.~Waterman, R.~Avižienis,
  J.~Wawrzynek, and K.~Asanović, ``{Chisel: Constructing Hardware in a Scala
  Embedded Language},'' in \emph{DAC}, 2012.

\bibitem{baer2}
J.-L. Baer and T.-F. Chen, ``{An Effective On-chip Preloading Scheme to Reduce
  Data Access Penalty},'' in \emph{SC}, 1991.

\bibitem{domino}
M.~Bakhshalipour, P.~Lotfi-Kamran, and H.~Sarbazi-Azad, ``{Domino Temporal Data
  Prefetcher},'' in \emph{HPCA}, 2018.

\bibitem{shadow-stacks}
A.~Baratloo, N.~Singh, and T.~Tsai, ``{Transparent Run-time Defense Against
  Stack Smashing Attacks},'' in \emph{ATEC}, 2000.

\bibitem{barber2019specshield}
K.~Barber, A.~Bacha, L.~Zhou, Y.~Zhang, and R.~Teodorescu, ``Specshield:
  Shielding speculative data from microarchitectural covert channels,'' in
  \emph{PACT}, 2019.

\bibitem{droplet19}
A.~Basak, S.~Li, X.~Hu, S.~M. Oh, X.~Xie, L.~Zhao, X.~Jiang, and Y.~Xie,
  ``{Analysis and Optimization of the Memory Hierarchy for Graph Processing
  Workloads},'' in \emph{HPCA}, 2019.

\bibitem{bekerman1999correlated}
M.~Bekerman, S.~Jourdan, R.~Ronen, G.~Kirshenboim, L.~Rappoport, A.~Yoaz, and
  U.~Weiser, ``{Correlated Load-address Predictors},'' in \emph{ISCA}, 1999.

\bibitem{pythiaMicro2021}
R.~Bera, K.~Kanellopoulos, A.~Nori, T.~Shahroodi, S.~Subramoney, and O.~Mutlu,
  ``{Pythia: A Customizable Hardware Prefetching Framework Using Online
  Reinforcement Learning},'' in \emph{MICRO}, 2021.

\bibitem{dspatch}
R.~Bera, A.~V. Nori, O.~Mutlu, and S.~Subramoney, ``{DSPatch: Dual Spatial
  Pattern Prefetcher},'' in \emph{MICRO}, 2019.

\bibitem{generatinghints-beyls-jsysarch05}
K.~Beyls and E.~H. D’Hollander, ``{Generating Cache Hints for Improved
  Program Efficiency},'' in \emph{JSA}, 2005.

\bibitem{ppf}
E.~Bhatia, G.~Chacon, S.~Pugsley, E.~Teran, P.~V. Gratz, and D.~A. Jim\'{e}nez,
  ``{Perceptron-Based Prefetch Filtering},'' in \emph{ISCA}, 2019.

\bibitem{jump-oriented-programming}
T.~Bletsch, X.~Jiang, V.~Freeh, and Z.~Liang, ``{Jump-oriented Programming: A
  New Class of Code-reuse Attack},'' in \emph{ASIA CCS}, 2011.

\bibitem{MI6}
T.~Bourgeat, I.~Lebedev, A.~Wright, S.~Zhang, Arvind, and S.~Devadas, ``{MI6:
  Secure Enclaves in a Speculative Out-of-Order Processor},'' in \emph{MICRO},
  2009.

\bibitem{pacman-brock-ismm13}
J.~Brock, X.~Gu, B.~Bao, and C.~Ding, ``{Pacman: Program-assisted Cache
  Management},'' in \emph{ISMM}, 2013.

\bibitem{carter1994hardware}
N.~P. Carter, S.~W. Keckler, and W.~J. Dally, ``{Hardware Support for Fast
  Capability-based Addressing},'' in \emph{ASPLOS}, 1994.

\bibitem{rop-wo-return}
S.~Checkoway, L.~Davi, A.~Dmitrienko, A.-R. Sadeghi, H.~Shacham, and
  M.~Winandy, ``{Return-oriented Programming Without Returns},'' in \emph{CCS},
  2010.

\bibitem{stride1}
T.-F. Chen and J.-L. Baer, ``{Effective Hardware-Based Data Prefetching for
  High-Performance Processors},'' in \emph{TC}, 1995.

\bibitem{CoopPref98}
{Chi-Keung Luk} and T.~C. {Mowry}, ``{Cooperative Prefetching: Compiler and
  Hardware Support for Effective Instruction Prefetching in Modern
  Processors},'' in \emph{MICRO}, 1998.

\bibitem{chilimbi2002dynamic}
T.~M. Chilimbi and M.~Hirzel, ``{Dynamic Hot Data Stream Prefetching for
  General-Purpose Programs},'' in \emph{PLDI}, 2002.

\bibitem{Sunder94}
T.~Chiueh, ``{Sunder: A Programmable Hardware Prefetch Architecture for
  Numerical Loops},'' in \emph{SC}, 1994.

\bibitem{chou2007low}
Y.~Chou, ``{Low-cost Epoch-based Correlation Prefetching for Commercial
  Applications},'' in \emph{MICRO}, 2007.

\bibitem{deputy}
J.~Condit, M.~Harren, Z.~Anderson, D.~Gay, and G.~C. Necula, ``{Dependent Types
  for Low-level Programming},'' in \emph{ESOP}, 2007.

\bibitem{cooksey2002stateless}
R.~Cooksey, S.~Jourdan, and D.~Grunwald, ``{A Stateless, Content-directed Data
  Prefetching Mechanism},'' in \emph{ASPLOS}, 2002.

\bibitem{canaries}
C.~Cowan, C.~Pu, D.~Maier, H.~Hintony, J.~Walpole, P.~Bakke, S.~Beattie,
  A.~Grier, P.~Wagle, and Q.~Zhang, ``{StackGuard: Automatic Adaptive Detection
  and Prevention of Buffer-overflow Attacks},'' in \emph{USENIX Security},
  1998.

\bibitem{minos}
J.~R. Crandall, S.~F. Wu, and F.~T. Chong, ``{Minos: Architectural Support for
  Protecting Control Data},'' in \emph{TACO}, 2006.

\bibitem{write-what-where}
\BIBentryALTinterwordspacing
{CWE MITRE}, ``{CWE-123: Write-what-where Condition},'' 2019. [Online].
  Available: \url{https://cwe.mitre.org/data/definitions/123.html}
\BIBentrySTDinterwordspacing

\bibitem{shadow-stacks-bad}
T.~H. Dang, P.~Maniatis, and D.~Wagner, ``{The Performance Cost of Shadow
  Stacks and Stack Canaries},'' in \emph{ASIA CCS}, 2015.

\bibitem{das2013a2c}
R.~Das, R.~Ausavarungnirun, O.~Mutlu, A.~Kumar, and M.~Azimi,
  ``{Application-to-core Mapping Policies to Reduce Memory System Interference
  in Multi-core Systems},'' in \emph{HPCA}, 2013.

\bibitem{trafficmanagement-dashti-asplos13}
M.~Dashti, A.~Fedorova, J.~Funston, F.~Gaud, R.~Lachaize, B.~Lepers, V.~Quéma,
  and M.~Roth, ``{Traffic Management: A Holistic Approach to Memory Placement
  on {NUMA} Systems},'' in \emph{ASPLOS}, 2013.

\bibitem{harmoni}
D.~Y. Deng and G.~E. Suh, ``{High-performance Parallel Accelerator for Flexible
  and Efficient Run-time Monitoring},'' in \emph{DSN}, 2012.

\bibitem{hardbound}
J.~Devietti, C.~Blundell, M.~M.~K. Martin, and S.~Zdancewic, ``{Hardbound:
  Architectural Support for Spatial Safety of the C Programming Language},'' in
  \emph{ASPLOS}, 2008.

\bibitem{software-defined-metadata}
U.~Dhawan, C.~Hritcu, R.~Rubin, N.~Vasilakis, S.~Chiricescu, J.~M. Smith, T.~F.
  Knight, Jr., B.~C. Pierce, and A.~DeHon, ``Architectural support for
  software-defined metadata processing,'' in \emph{ASPLOS}, 2015.

\bibitem{pump}
U.~Dhawan, N.~Vasilakis, R.~Rubin, S.~Chiricescu, J.~M. Smith, T.~F. Knight,
  Jr., B.~C. Pierce, and A.~DeHon, ``{PUMP: A Programmable Unit for Metadata
  Processing},'' in \emph{HASP}, 2014.

\bibitem{backwards-compatible-bounds-checking}
D.~Dhurjati and V.~Adve, ``{Backwards-compatible Array Bounds Checking for C
  with Very Low Overhead},'' in \emph{ICSE}, 2006.

\bibitem{datatiering-dulloor-eurosys16}
S.~R. Dulloor, A.~Roy, Z.~Zhao, N.~Sundaram, N.~Satish, R.~Sankaran,
  J.~Jackson, and K.~Schwan, ``{Data Tiering in Heterogeneous Memory
  Systems},'' in \emph{EuroSys}, 2016.

\bibitem{ebrahimi-10}
E.~Ebrahimi, C.~J. Lee, O.~Mutlu, and Y.~N. Patt, ``{Fairness via Source
  Throttling: A Configurable and High-performance Fairness Substrate for
  Multi-core Memory Systems},'' in \emph{ASPLOS}, 2010.

\bibitem{ebrahimi2009techniques}
E.~Ebrahimi, O.~Mutlu, and Y.~N. Patt, ``{Techniques for Bandwidth-efficient
  Prefetching of Linked Data Structures in Hybrid Prefetching Systems},'' in
  \emph{HPCA}, 2009.

\bibitem{checked-c}
A.~S. Elliott, A.~Ruef, M.~Hicks, and D.~Tarditi, ``{Checked C: Making C Safe
  by Extension},'' in \emph{SecDev}, 2018.

\bibitem{ferdman2007last}
M.~Ferdman and B.~Falsafi, ``{Last-touch Correlated Data Streaming},'' in
  \emph{ISPASS}, 2007.

\bibitem{sms_mod}
M.~Ferdman, S.~Somogyi, and B.~Falsafi, ``{Spatial Memory Streaming with
  Rotated Patterns},'' in \emph{In 1st JILP Data Prefetching Championship},
  2009.

\bibitem{tagged-feustel-taco73}
E.~A. Feustel, ``{On The Advantages of Tagged Architecture},'' in \emph{TC},
  1973.

\bibitem{stride_vector}
J.~W.~C. Fu and J.~H. Patel, ``{Data Prefetching in Multiprocessor Vector Cache
  Memories},'' in \emph{ISCA}, 1991.

\bibitem{stride}
J.~W.~C. Fu, J.~H. Patel, and B.~L. Janssens, ``{Stride Directed Prefetching in
  Scalar Processors},'' in \emph{MICRO}, 1992.

\bibitem{buddy-allocator}
\BIBentryALTinterwordspacing
M.~Gorman, ``{Physical Page Allocation},'' 2007. [Online]. Available:
  \url{https://www.kernel.org/doc/gorman/html/understand/understand009.html}
\BIBentrySTDinterwordspacing

\bibitem{metasysGithub}
\BIBentryALTinterwordspacing
S.~R. Group, ``{MetaSys},'' 2021. [Online]. Available:
  \url{https://github.com/CMU-SAFARI/MetaSys}
\BIBentrySTDinterwordspacing

\bibitem{popt-gu-lcpc08}
X.~Gu, T.~Bai, Y.~Gao, C.~Zhang, R.~Archambault, and C.~Ding, ``{P-OPT:
  Program-Directed Optimal Cache Management},'' in \emph{LCPC}, 2008.

\bibitem{hashemi2018learning}
M.~Hashemi, K.~Swersky, J.~Smith, G.~Ayers, H.~Litz, J.~Chang, C.~Kozyrakis,
  and P.~Ranganathan, ``{Learning Memory Access Patterns},'' in \emph{ICML},
  2018.

\bibitem{purify}
R.~Hastings and B.~Joyce, ``{Purify: Fast Detection of Memory Leaks and Access
  Errors},'' in \emph{USENIX}, 1991.

\bibitem{hu2003tcp}
Z.~Hu, M.~Martonosi, and S.~Kaxiras, ``{TCP: Tag Correlating Prefetchers},'' in
  \emph{HPCA}, 2003.

\bibitem{labeled-yu}
B.~Huang, X.~Jin, H.~Wang, Y.~Zhou, Z.~Chang, Y.~Cao, and Y.~Bao, ``{Labeled
  {RISC-V}: A New Perspective on Software-Defined Architecture},'' in
  \emph{CARVV}, 2017.

\bibitem{cet}
\BIBentryALTinterwordspacing
Intel, ``{Control-flow Enforcement Technology Specification},'' 2019. [Online].
  Available:
  \url{www.intel.com/content/dam/www/public/us/en/documents/white-papers/virtualization-enabling-
  intel-virtualization-technology-features-and-benefits-paper.pdf}
\BIBentrySTDinterwordspacing

\bibitem{ampm}
Y.~Ishii, M.~Inaba, and K.~Hiraki, ``{Access Map Pattern Matching for Data
  Cache Prefetch},'' in \emph{ISC}, 2009.

\bibitem{isb}
A.~Jain and C.~Lin, ``{Linearizing Irregular Memory Accesses for Improved
  Correlated Prefetching},'' in \emph{MICRO}, 2013.

\bibitem{swcache-jain-iccad01}
P.~Jain, S.~Devadas, D.~Engels, and L.~Rudolph, ``{Software-Assisted Cache
  Replacement Mechanisms for Embedded Systems},'' in \emph{ICCAD}, 2001.

\bibitem{cyclone}
T.~Jim, J.~G. Morrisett, D.~Grossman, M.~W. Hicks, J.~Cheney, and Y.~Wang,
  ``{Cyclone: A Safe Dialect of C},'' in \emph{ATEC}, 2002.

\bibitem{efficient-tagged-memory}
A.~{Joannou}, J.~{Woodruff}, R.~{Kovacsics}, S.~W. {Moore}, A.~{Bradbury},
  H.~{Xia}, R.~N.~M. {Watson}, D.~{Chisnall}, M.~{Roe}, B.~{Davis},
  E.~{Napierala}, J.~{Baldwin}, K.~{Gudka}, P.~G. {Neumann}, A.~{Mazzinghi},
  A.~{Richardson}, S.~{Son}, and A.~T. {Markettos}, ``{Efficient Tagged
  Memory},'' in \emph{ICCD}, 2017.

\bibitem{joao2009flexible}
J.~A. Joao, O.~Mutlu, and Y.~N. Patt, ``{Flexible Reference-Counting-Based
  Hardware Acceleration for Garbage Collection},'' in \emph{ISCA}, 2009.

\bibitem{markov}
D.~Joseph and D.~Grunwald, ``{Prefetching using Markov predictors},'' in
  \emph{ISCA}, 1997.

\bibitem{jouppi_prefetch}
N.~P. Jouppi, ``{Improving Direct-mapped Cache Performance by the Addition of a
  Small Fully-associative Cache and Prefetch Buffers},'' in \emph{ISCA}, 1990.

\bibitem{smash}
K.~Kanellopoulos, N.~Vijaykumar, C.~Giannoula, R.~Azizi, S.~Koppula, N.~M.
  Ghiasi, T.~Shahroodi, J.~G. Luna, and O.~Mutlu, ``{SMASH: Co-Designing
  Software Compression and Hardware-Accelerated Indexing for Efficient Sparse
  Matrix Operations},'' in \emph{MICRO}, 2019.

\bibitem{karlsson2000prefetching}
M.~Karlsson, F.~Dahlgren, and P.~Stenstrom, ``{A Prefetching Technique for
  Irregular Accesses to Linked Data Structures},'' in \emph{HPCA}, 2000.

\bibitem{khasawneh2019safespec}
K.~N. Khasawneh, E.~M. Koruyeh, C.~Song, D.~Evtyushkin, D.~Ponomarev, and
  N.~Abu-Ghazaleh, ``{SafeSpec: Banishing the Spectre of a Meltdown with
  Leakage-Free Speculation},'' in \emph{DAC}, 2019.

\bibitem{spp}
J.~Kim, S.~H. Pugsley, P.~V. Gratz, A.~Reddy, C.~Wilkerson, and Z.~Chishti,
  ``{Path Confidence based Lookahead Prefetching},'' in \emph{MICRO}, 2016.

\bibitem{tcm-kim-micro10}
Y.~Kim, M.~Papamichael, O.~Mutlu, and M.~Harchol-Balter, ``{Thread Cluster
  Memory Scheduling: Exploiting Differences in Memory Access Behavior},'' in
  \emph{MICRO}, 2010.

\bibitem{buddy-original}
K.~C. Knowlton, ``{A Fast Storage Allocator},'' in \emph{CACM}, 1965.

\bibitem{timing-attacks}
P.~C. Kocher, ``{Timing Attacks on Implementations of Diffie-Hellman, RSA, DSS,
  and Other Systems},'' in \emph{CRYPTO}, 1996.

\bibitem{ext-mondrian}
\BIBentryALTinterwordspacing
C.~Kolbitsch, C.~Kruegel, and E.~Kirda, ``Extending mondrian memory
  protection,'' 2011. [Online]. Available:
  \url{https://www.sto.nato.int/publications/STO\%20Meeting\%20Proceedings/RTO-MP-IST-091/MP-IST-091-10.pdf}
\BIBentrySTDinterwordspacing

\bibitem{dol}
S.~Kondguli and M.~Huang, ``{Division of labor: A more effective approach to
  prefetching},'' in \emph{ISCA}, 2018.

\bibitem{eden}
S.~Koppula, L.~Orosa, A.~G. Ya\u{g}l\i{}k\c{c}\i{}, R.~Azizi, T.~Shahroodi,
  K.~Kanellopoulos, and O.~Mutlu, ``{EDEN: Enabling Energy-Efficient,
  High-Performance Deep Neural Network Inference Using Approximate DRAM},'' in
  \emph{MICRO}, 2019.

\bibitem{footprint}
S.~Kumar and C.~Wilkerson, ``{Exploiting Spatial Locality in Data Caches using
  Spatial Footprints},'' in \emph{ISCA}, 1998.

\bibitem{cpi}
V.~Kuznetsov, L.~Szekeres, M.~Payer, G.~Candea, R.~Sekar, and D.~Song,
  ``{Code-Pointer Integrity},'' in \emph{OSDI}, 2014.

\bibitem{lowfat-pointers}
A.~Kwon, U.~Dhawan, J.~M. Smith, T.~F. Knight, Jr., and A.~DeHon, ``{Low-fat
  Pointers: Compact Encoding and Efficient Gate-level Implementation of Fat
  Pointers for Spatial Safety and Capability-based Security},'' in \emph{CCS},
  2013.

\bibitem{levy2014capability}
H.~M. Levy, \emph{{Capability-based Computer Systems}}, 1984.

\bibitem{pac-it-up}
H.~Liljestrand, T.~Nyman, K.~Wang, C.~C. Perez, J.-E. Ekberg, and N.~Asokan,
  ``{PAC it up: Towards Pointer Integrity using ARM Pointer Authentication},''
  in \emph{USENIX Security}, 2018.

\bibitem{flikker-liu-asplos11}
S.~Liu, K.~Pattabiraman, T.~Moscibroda, and B.~G. Zorn, ``{Flikker: Saving
  {DRAM} Refresh-power Through Critical Data Partitioning},'' in \emph{ASPLOS},
  2011.

\bibitem{avio}
S.~Lu, J.~Tucek, F.~Qin, and Y.~Zhou, ``{AVIO: Detecting Atomicity Violations
  via Access Interleaving Invariants},'' in \emph{ASPLOS}, 2006.

\bibitem{colorsafe}
B.~Lucia, L.~Ceze, and K.~Strauss, ``{ColorSafe: Architectural Support for
  Debugging and Dynamically Avoiding Multi-variable Atomicity Violations},'' in
  \emph{ISCA}, 2010.

\bibitem{conflictexceptions}
B.~Lucia, L.~Ceze, K.~Strauss, S.~Qadeer, and H.-J. Boehm, ``{Conflict
  Exceptions: Simplifying Concurrent Language Semantics with Precise Hardware
  Exceptions for Data-races},'' in \emph{ISCA}, 2010.

\bibitem{lucia2008atom}
B.~Lucia, J.~Devietti, K.~Strauss, and L.~Ceze, ``{Atom-aid: Detecting and
  Surviving Atomicity Violations},'' in \emph{ISCA}, 2008.

\bibitem{characterizing-luo-dsn14}
Y.~Luo, S.~Govindan, B.~Sharma, M.~Santaniello, J.~Meza, A.~Kansal, J.~Liu,
  B.~Khessib, K.~Vaid, and O.~Mutlu, ``{Characterizing Application Memory Error
  Vulnerability to Optimize Datacenter Cost via Heterogeneous-reliability
  Memory},'' in \emph{DSN}, 2014.

\bibitem{pard-ma-asplos15}
J.~Ma, X.~Sui, N.~Sun, Y.~Li, Z.~Yu, B.~Huang, T.~Xu, Z.~Yao, Y.~Chen, H.~Wang,
  L.~Zhang, and Y.~Bao, ``{Supporting Differentiated Services in Computers via
  Programmable Architecture for Resourcing-on-Demand ({PARD})},'' in
  \emph{ASPLOS}, 2015.

\bibitem{maas2016grail}
M.~Maas, K.~Asanovic, and J.~Kubiatowicz, ``{Grail Quest: A New Proposal for
  Hardware-assisted Garbage Collection},'' in \emph{Workshop on Architectures
  and Systems for Big Data}, 2016.

\bibitem{maas2018hardware}
M.~Maas, K.~Asanovi{\'c}, and J.~Kubiatowicz, ``{A Hardware Accelerator for
  Tracing Garbage Collection},'' in \emph{ISCA}, 2018.

\bibitem{radar-manivannan-hpca16}
M.~{Manivannan}, V.~{Papaefstathiou}, M.~{Pericas}, and P.~{Stenstrom},
  ``{RADAR: Runtime-Assisted Dead Region Management for Last-Level Caches},''
  in \emph{HPCA}, 2016.

\bibitem{CCFI}
A.~J. Mashtizadeh, A.~Bittau, D.~Boneh, and D.~Mazi\`{e}res, ``{CCFI:
  Cryptographically Enforced Control Flow Integrity},'' in \emph{CCS}, 2015.

\bibitem{shaktiT}
A.~Menon, S.~Murugan, C.~Rebeiro, N.~Gala, and K.~Veezhinathan, ``{Shakti-{T}:
  A {RISC-V} Processor with Light Weight Security Extensions},'' in
  \emph{HASP}, 2017.

\bibitem{bop}
P.~Michaud, ``{Best-offset Hardware Prefetching},'' in \emph{HPCA}, 2016.

\bibitem{nescheck}
D.~Midi, M.~Payer, and E.~Bertino, ``{Memory Safety for Embedded Devices with
  nesCheck},'' in \emph{ASIA CCS}, 2017.

\bibitem{OCFI}
V.~Mohan, P.~Larsen, S.~Brunthaler, K.~W. Hamlen, and M.~Franz, ``{Opaque
  Control-Flow Integrity},'' in \emph{NDSS}, 2015.

\bibitem{cfs}
\BIBentryALTinterwordspacing
I.~Molnar, ``{Modular Scheduler Core and Completely Fair Scheduler},'' 2007.
  [Online]. Available:
  \url{https://web.archive.org/web/20070419102054/http://kerneltrap.org/node/8059}
\BIBentrySTDinterwordspacing

\bibitem{ets18}
A.~Mukkara, N.~Beckmann, M.~Abeydeera, X.~Ma, and D.~Sanchez, ``{Exploiting
  Locality in Graph Analytics Through Hardware-accelerated Traversal
  Scheduling},'' in \emph{MICRO}, 2018.

\bibitem{whirlpool-mukkara-asplos16}
A.~Mukkara, N.~Beckmann, and D.~Sanchez, ``{Whirlpool: Improving Dynamic Cache
  Management with Static Data Classification},'' in \emph{ASPLOS}, 2016.

\bibitem{mcp-muralidhara-micro11}
Muralidhara, L.~Subramanian, O.~Mutlu, M.~Kandemir, and T.~Moscibroda,
  ``{Reducing Memory Interference in Multicore Systems via Application-aware
  Memory Channel Partitioning},'' in \emph{MICRO}, 2011.

\bibitem{mutlu2007stall}
O.~Mutlu and T.~Moscibroda, ``{Stall-Time Fair Memory Access Scheduling for
  Chip Multiprocessors},'' in \emph{MICRO}, 2007.

\bibitem{parbs-mutlu-isca08}
O.~Mutlu and T.~Moscibroda, ``{Parallelism-Aware Batch Scheduling: Enhancing
  both Performance and Fairness of Shared {DRAM} Systems},'' in \emph{ISCA},
  2008.

\bibitem{atomtracker}
A.~{Muzahid}, N.~{Otsuki}, and J.~{Torrellas}, ``{AtomTracker: A Comprehensive
  Approach to Atomic Region Inference and Violation Detection},'' in
  \emph{MICRO}, 2010.

\bibitem{watchdog}
S.~Nagarakatte, M.~M.~K. Martin, and S.~Zdancewic, ``{Watchdog: Hardware for
  Safe and Secure Manual Memory Management and Full Memory Safety},'' in
  \emph{ISCA}, 2012.

\bibitem{watchdog-lite}
S.~Nagarakatte, M.~M.~K. Martin, and S.~Zdancewic, ``{WatchdogLite:
  Hardware-Accelerated Compiler-Based Pointer Checking},'' in \emph{CGO}, 2014.

\bibitem{softbound}
S.~Nagarakatte, J.~Zhao, M.~M. Martin, and S.~Zdancewic, ``{SoftBound: Highly
  Compatible and Complete Spatial Memory Safety for C},'' in \emph{PLDI}, 2009.

\bibitem{ccured}
G.~C. Necula, J.~Condit, M.~Harren, S.~McPeak, and W.~Weimer, ``{CCured:
  Type-safe Retrofitting of Legacy Software},'' in \emph{TOPLAS}, 2005.

\bibitem{PrefEdge2014}
K.~Nilakant, V.~Dalibard, A.~Roy, and E.~Yoneki, ``{PrefEdge: SSD Prefetcher
  for Large-Scale Graph Traversal},'' in \emph{SYSTOR}, 2014.

\bibitem{mpx-explained}
O.~Oleksenko, D.~Kuvaiskii, P.~Bhatotia, P.~Felber, and C.~Fetzer, ``{Intel
  {MPX} Explained: A Cross-layer Analysis of the {Intel} {MPX} System Stack},''
  in \emph{SIGMETRICS}, 2018.

\bibitem{ipcp}
S.~Pakalapati and B.~Panda, ``{Bouquet of Instruction Pointers: Instruction
  Pointer Classifier-based Spatial Hardware Prefetching},'' in \emph{ISCA},
  2020.

\bibitem{runtimellc-pan-sc15}
A.~Pan and V.~S. Pai, ``{Runtime-driven Shared Last-level Cache Management for
  Task-parallel Programs},'' in \emph{SC}, 2015.

\bibitem{prefetchtasklifetimes-papaefstathiou-ics13}
V.~Papaefstathiou, M.~G. Katevenis, D.~S. Nikolopoulos, and D.~Pnevmatikatos,
  ``{Prefetching and Cache Management Using Task Lifetimes},'' in \emph{ICS},
  2013.

\bibitem{irregularities-park-asplos13}
H.~Park, S.~Baek, J.~Choi, D.~Lee, and S.~H. Noh, ``{Regularities Considered
  Harmful: Forcing Randomness to Memory Accesses to Reduce Row Buffer Conflicts
  for Multi-core, Multi-bank Systems},'' in \emph{ASPLOS}, 2013.

\bibitem{shadow-processing}
H.~Patil and C.~N. Fischer, ``{Efficient Run-time Monitoring Using Shadow
  Processing},'' in \emph{AADEBUG}, 1995.

\bibitem{peled_rl}
L.~{Peled}, S.~{Mannor}, U.~{Weiser}, and Y.~{Etsion}, ``{Semantic Locality and
  Context-based Prefetching using Reinforcement Learning},'' in \emph{ISCA},
  2015.

\bibitem{PrefReinforce2015}
L.~Peled, S.~Mannor, U.~Weiser, and Y.~Etsion, ``{Semantic Locality and
  Context-based Prefetching Using Reinforcement Learning},'' in \emph{ISCA},
  2015.

\bibitem{peled2018neural}
L.~Peled, U.~Weiser, and Y.~Etsion, ``{A Neural Network Memory Prefetcher Using
  Semantic Locality},'' arXiv:1804.00478, 2018.

\bibitem{iwatcher}
{Pin Zhou}, {Feng Qin}, {Wei Liu}, {Yuanyuan Zhou}, and J.~{Torrellas},
  ``{iWatcher: Efficient Architectural Support for Software Debugging},'' in
  \emph{ISCA}, 2004.

\bibitem{polybench}
\BIBentryALTinterwordspacing
L.~Pouchet, ``Polybench: The polyhedral benchmark suite,'' 2015. [Online].
  Available: \url{http://web.cse.ohio-state.edu/~pouchet.2/software/polybench/}
\BIBentrySTDinterwordspacing

\bibitem{sandbox}
S.~H. Pugsley, Z.~Chishti, C.~Wilkerson, P.-f. Chuang, R.~L. Scott, A.~Jaleel,
  S.-L. Lu, K.~Chow, and R.~Balasubramonian, ``{Sandbox Prefetching: Safe
  Run-time Evaluation of Aggressive Prefetchers},'' in \emph{HPCA}, 2014.

\bibitem{compilerpartitioned-ravindran-lctes07}
R.~Ravindran, M.~Chu, and S.~Mahlke, ``{Compiler-managed Partitioned Data
  Caches for Low Power},'' in \emph{LCTES}, 2007.

\bibitem{tempest-typhoon}
S.~K. Reinhardt, J.~R. Larus, and D.~A. Wood, ``{Tempest and Typhoon:
  User-level Shared Memory},'' in \emph{ISCA}, 1994.

\bibitem{riscv-pk}
\BIBentryALTinterwordspacing
RISC-V, ``{RISC-V} proxy kernel,'' 2019. [Online]. Available:
  \url{https://github.com/riscv/riscv-pk}
\BIBentrySTDinterwordspacing

\bibitem{protecting-stack}
N.~Roessler and A.~Dehon, ``{Protecting the Stack with Metadata Policies and
  Tagged Hardware},'' in \emph{SP}, 2018.

\bibitem{olden}
A.~Rogers, M.~C. Carlisle, J.~H. Reppy, and L.~J. Hendren, ``{Supporting
  Dynamic Data Structures on Distributed-Memory Machines},'' in \emph{TOPLAS},
  1995.

\bibitem{DepBasedPref98}
A.~Roth, A.~Moshovos, and G.~S. Sohi, ``{Dependence Based Prefetching for
  Linked Data Structures},'' in \emph{ASPLOS}, 1998.

\bibitem{roth1999Effective}
A.~Roth and G.~S. Sohi, ``{Effective Jump-Pointer Prefetching for Linked Data
  Structures},'' in \emph{ISCA}, 1999.

\bibitem{practical-overflow-detector}
O.~Ruwase and M.~S. Lam, ``{A Practical Dynamic Buffer Overflow Detector},'' in
  \emph{NDSS}, 2004.

\bibitem{cleanupspec}
G.~Saileshwar and M.~K. Qureshi, ``{CleanupSpec: An “Undo” Approach to Safe
  Speculation},'' in \emph{MICRO}, 2019.

\bibitem{approximate-sampson-tocs14}
A.~Sampson, J.~Nelson, K.~Strauss, and L.~Ceze, ``{Approximate Storage in
  Solid-state Memories},'' in \emph{TOCS}, 2014.

\bibitem{cooperativescrubbing-sartor-pact14}
J.~B. Sartor, W.~Heirman, S.~M. Blackburn, L.~Eeckhout, and K.~S. McKinley,
  ``{Cooperative Cache Scrubbing},'' in \emph{PACT}, 2014.

\bibitem{keepme-sartor-interact05}
J.~B. Sartor, S.~Venkiteswaran, K.~S. McKinley, and Z.~Wang, ``{Cooperative
  Caching with Keep-me and Evict-me},'' in \emph{INTERACT}, 2005.

\bibitem{seshadri2015page}
V.~Seshadri, G.~Pekhimenko, O.~Ruwase, O.~Mutlu, P.~B. Gibbons, M.~A. Kozuch,
  T.~C. Mowry, and T.~Chilimbi, ``{Page Overlays: An Enhanced Virtual Memory
  Framework to Enable Fine-Grained Memory Management},'' in \emph{ISCA}, 2015.

\bibitem{mlop}
M.~Shakerinava, M.~Bakhshalipour, P.~Lotfi-Kamran, and H.~Sarbazi-Azad,
  ``{Multi-Lookahead Offset Prefetching},'' 3rd Data Prefetching Championship,
  2019.

\bibitem{vldp}
M.~Shevgoor, S.~Koladiya, R.~Balasubramonian, C.~Wilkerson, S.~H. Pugsley, and
  Z.~Chishti, ``Efficiently prefetching complex address patterns,'' in
  \emph{MICRO}, 2015.

\bibitem{shineural}
Z.~Shi, A.~Jain, K.~Swersky, M.~Hashemi, P.~Ranganathan, and C.~Lin, ``{A
  Neural Hierarchical Sequence Model for Irregular Data Prefetching},'' in
  \emph{ML For Systems Workshop, NeurIPS}, 2019.

\bibitem{shineural_asplos}
Z.~Shi, A.~Jain, K.~Swersky, M.~Hashemi, P.~Ranganathan, and C.~Lin, ``{A
  Hierarchical Neural Model of Data Prefetching},'' in \emph{ASPLOS}, 2021.

\bibitem{shi2019learning}
Z.~Shi, K.~Swersky, D.~Tarlow, P.~Ranganathan, and M.~Hashemi, ``{Learning
  Execution Through Neural Code Fusion},'' arXiv:1906.07181, 2019.

\bibitem{intel-22nm-measurement}
\BIBentryALTinterwordspacing
A.~L. Shimpi, ``{Dual Core/GT2 Ivy Bridge Die Measured: 121mm2},'' 2012.
  [Online]. Available:
  \url{https://www.anandtech.com/show/5875/dual-coregt2-ivy-bridge-die-measured-121mm2}
\BIBentrySTDinterwordspacing

\bibitem{ligra}
J.~Shun and G.~E. Blelloch, ``{Ligra: A Lightweight Graph Processing Framework
  for Shared Memory},'' in \emph{PPoPP}, 2013.

\bibitem{memsafe}
M.~S. Simpson and R.~K. Barua, ``{MemSafe: Ensuring the Spatial and Temporal
  Memory Safety of C at Runtime},'' in \emph{SCAM}, 2010.

\bibitem{somogyi_stems}
S.~Somogyi, T.~F. Wenisch, A.~Ailamaki, and B.~Falsafi, ``{Spatio-Temporal
  Memory Streaming},'' in \emph{ISCA}, 2009.

\bibitem{sms}
S.~Somogyi, T.~F. Wenisch, A.~Ailamaki, B.~Falsafi, and A.~Moshovos, ``{Spatial
  Memory Streaming},'' in \emph{ISCA}, 2006.

\bibitem{hdfi}
C.~Song, H.~Moon, M.~Alam, I.~Yun, B.~Lee, T.~Kim, W.~Lee, and Y.~Paek,
  ``{HDFI: Hardware-Assisted Data-Flow Isolation},'' in \emph{SP}, 2016.

\bibitem{fdp}
S.~Srinath, O.~Mutlu, H.~Kim, and Y.~N. Patt, ``{Feedback Directed Prefetching:
  Improving the Performance and Bandwidth-efficiency of Hardware
  Prefetchers},'' in \emph{HPCA}, 2007.

\bibitem{breaking-memory-secrecy}
R.~Strackx, Y.~Younan, P.~Philippaerts, F.~Piessens, S.~Lachmund, and
  T.~Walter, ``{Breaking the Memory Secrecy Assumption},'' in \emph{EUROSEC},
  2009.

\bibitem{subramanian2015asm}
L.~Subramanian, V.~Seshadri, A.~Ghosh, S.~Khan, and O.~Mutlu, ``{The
  Application Slowdown Model: Quantifying and Controlling the Impact of
  Inter-Application Interference at Shared Caches and Main Memory},'' in
  \emph{MICRO}, 2015.

\bibitem{subramanian2013mise}
L.~Subramanian, V.~Seshadri, Y.~Kim, B.~Jaiyen, and O.~Mutlu, ``{MISE:
  Providing Performance Predictability and Improving Fairness in Shared Main
  Memory Systems},'' in \emph{HPCA}, 2013.

\bibitem{spe}
G.~E. Suh, J.~W. Lee, D.~Zhang, and S.~Devadas, ``{Secure Program Execution via
  Dynamic Information Flow Tracking},'' in \emph{ASPLOS}, 2004.

\bibitem{synopsys}
\BIBentryALTinterwordspacing
Synopsys, ``{Synopsys Design Compiler},'' 2021. [Online]. Available:
  \url{https://www.synopsys.com/support/training/rtl-synthesis/design-compiler-rtl-synthesis.html}
\BIBentrySTDinterwordspacing

\bibitem{SoK}
L.~Szekeres, M.~Payer, T.~Wei, and D.~Song, ``{SoK: Eternal War in Memory},''
  in \emph{SP}, 2013.

\bibitem{prodigy-hpca21}
N.~Talati, K.~May, A.~Behroozi, Y.~Yang, K.~Kaszyk, C.~Vasiladiotis, T.~Verma,
  L.~Li, B.~Nguyen, J.~Sun, J.~M. Morton, A.~Ahmadi, T.~Austin, M.~O’Boyle,
  S.~Mahlke, T.~Mudge, and R.~Dreslinski, ``{Prodigy: Improving the Memory
  Latency of Data-Indirect Irregular Workloads Using Hardware-Software
  Co-Design},'' in \emph{HPCA}, 2021.

\bibitem{return-into-libc-expressiveness}
M.~Tran, M.~Etheridge, T.~Bletsch, X.~Jiang, V.~Freeh, and P.~Ning, ``{On the
  Expressiveness of Return-into-libc Attacks},'' in \emph{RAID}, 2011.

\bibitem{modified-tyson-micro95}
G.~Tyson, M.~Farrens, J.~Matthews, and A.~R. Pleszkun, ``{A Modified Approach
  to Data Cache Management},'' in \emph{MICRO}, 1995.

\bibitem{execshield}
\BIBentryALTinterwordspacing
A.~van~de Ven, ``{New Security Enhancements in Red Hat Enterprise Linux: v.3
  update 3},'' 2004. [Online]. Available:
  \url{https://static.redhat.com/legacy/f/pdf/rhel/WHP0006US_Execshield.pdf}
\BIBentrySTDinterwordspacing

\bibitem{memtracker}
G.~Venkataramani, B.~Roemer, Y.~Solihin, and M.~Prvulovic, ``{MemTracker:
  Efficient and Programmable Support for Memory Access Monitoring and
  Debugging},'' in \emph{HPCA}, 2007.

\bibitem{nanditaThesis}
N.~Vijaykumar, ``{Enhancing Programmability, Portability, and Performance with
  Rich Cross-Layer Abstractions},'' Ph.D. dissertation, 2019.

\bibitem{ldesc}
N.~Vijaykumar, E.~Ebrahimi, K.~Hsieh, P.~B. Gibbons, and O.~Mutlu, ``{The
  Locality Descriptor: A Holistic Cross-Layer Abstraction to Express Data
  Locality in GPUs},'' in \emph{ISCA}, 2018.

\bibitem{xmem}
N.~Vijaykumar, A.~Jain, D.~Majumdar, K.~Hsieh, G.~Pekhimenko, E.~Ebrahimi,
  N.~Hajinazar, P.~B. Gibbons, and O.~Mutlu, ``{A Case for Richer Cross-layer
  Abstractions: Bridging the Semantic Gap with Expressive Memory},'' in
  \emph{ISCA}, 2018.

\bibitem{NowatzkiISCA19}
Z.~Wang and T.~Nowatzki, ``{Stream-based Memory Access Specialization for
  General Purpose Processors},'' in \emph{ISCA}, 2019.

\bibitem{PrefCache2004}
Z.~Wang, K.~S. Mckinley, and D.~Burger, ``{Combining Cooperative
  Software/Hardware Prefetching and Cache Replacement},'' in \emph{IBM Austin
  CAS Center for Advanced Studies Conference}, 2004.

\bibitem{evictme-wang-pact02}
Z.~Wang, K.~S. McKinley, A.~L. Rosenberg, and C.~C. Weems, ``{Using the
  Compiler to Improve Cache Replacement Decisions},'' in \emph{PACT}, 2002.

\bibitem{watsoncapability}
R.~N.~M. Watson, J.~Anderson, B.~Laurie, and K.~Kennaway, ``{Capsicum:
  Practical Capabilities for UNIX},'' in \emph{USENIX Security}, 2010.

\bibitem{stems}
T.~F. Wenisch, M.~Ferdman, A.~Ailamaki, B.~Falsafi, and A.~Moshovos,
  ``{Practical Off-chip Meta-data for Temporal Memory Streaming},'' in
  \emph{HPCA}, 2009.

\bibitem{wenisch2010making}
T.~F. Wenisch, M.~Ferdman, A.~Ailamaki, B.~Falsafi, and A.~Moshovos, ``{Making
  Address-correlated Prefetching Practical},'' in \emph{IEEE Micro}, 2010.

\bibitem{wenisch2005temporal}
T.~F. Wenisch, S.~Somogyi, N.~Hardavellas, J.~Kim, A.~Ailamaki, and B.~Falsafi,
  ``{Temporal streaming of shared memory},'' in \emph{ISCA}, 2005.

\bibitem{legba}
A.~Wiggins, S.~Winwood, H.~Tuch, and G.~Heiser, ``{Legba: Fast Hardware Support
  for Fine-Grained Protection},'' in \emph{ASIA CCS}, 2003.

\bibitem{mondrian-witchel-asplos02}
E.~Witchel, J.~Cates, and K.~Asanovi\'{c}, ``{Mondrian Memory Protection},'' in
  \emph{ASPLOS}, 2002.

\bibitem{cheri}
J.~Woodruff, R.~N. Watson, D.~Chisnall, S.~W. Moore, J.~Anderson, B.~Davis,
  B.~Laurie, P.~G. Neumann, R.~Norton, and M.~Roe, ``{The {CHERI} Capability
  Model: Revisiting {RISC} in an Age of Risk},'' in \emph{ISCA}, 2014.

\bibitem{triage}
H.~Wu, K.~Nathella, J.~Pusdesris, D.~Sunwoo, A.~Jain, and C.~Lin, ``{Temporal
  Prefetching Without the Off-Chip Metadata},'' in \emph{MICRO}, 2019.

\bibitem{misb}
H.~Wu, K.~Nathella, D.~Sunwoo, A.~Jain, and C.~Lin, ``{Efficient Metadata
  Management for Irregular Data Prefetching},'' in \emph{ISCA}, 2019.

\bibitem{efficient-backwards-transform}
W.~Xu, D.~C. DuVarney, and R.~Sekar, ``{An Efficient and Backwards-compatible
  Transformation to Ensure Memory Safety of C Programs},'' in \emph{SIGSOFT},
  2004.

\bibitem{invisispec}
M.~Yan, J.~Choi, D.~Skarlatos, A.~Morrison, C.~W. Fletcher, and J.~Torrellas,
  ``{InvisiSpec: Making Speculative Execution Invisible in the Cache
  Hierarchy},'' in \emph{MICRO}, 2018.

\bibitem{compilerassisted-yang-lcpc04}
H.~Yang, R.~Govindarajan, G.~R. Gao, and Z.~Hu, ``{Compiler-Assisted Cache
  Replacement: Problem Formulation and Performance Evaluation},'' in
  \emph{LCPC}, 2003.

\bibitem{protecting-c-programs}
S.~H. Yong and S.~Horwitz, ``{Protecting C Programs from Attacks via Invalid
  Pointer Dereferences},'' in \emph{ESEC}, 2003.

\bibitem{imp15}
X.~Yu, C.~J. Hughes, N.~Satish, and S.~Devadas, ``{IMP: Indirect Memory
  Prefetcher},'' in \emph{MICRO}, 2015.

\bibitem{hardware-zeldovich-osdi08}
N.~Zeldovich, H.~Kannan, M.~Dalton, and C.~Kozyrakis, ``Hardware enforcement of
  application security policies using tagged memory,'' in \emph{OSDI}, 2008.

\bibitem{hardware-enforcement}
N.~Zeldovich, H.~Kannan, M.~Dalton, and C.~Kozyrakis, ``{Hardware Enforcement
  of Application Security Policies Using Tagged Memory},'' in \emph{OSDI},
  2008.

\bibitem{zeng2017long}
Y.~Zeng and X.~Guo, ``{Long Short Term Memory based Hardware Prefetcher: A Case
  Study},'' in \emph{MEMSYS}, 2017.

\bibitem{minnow}
D.~Zhang, X.~Ma, M.~Thomson, and D.~Chiou, ``{Minnow: Lightweight Offload
  Engines for Worklist Management and Worklist-Directed Prefetching},'' in
  \emph{ASPLOS}, 2018.

\bibitem{raguard}
J.~Zhang, R.~Hou, J.~Fan, K.~Liu, L.~Zhang, and S.~A. McKee, ``{RAGuard: A
  Hardware Based Mechanism for Backward-Edge Control-Flow Integrity},'' in
  \emph{CF}, 2017.

\bibitem{GuidedRegionPref2003}
{Zhenlin Wang}, D.~{Burger}, K.~S. {McKinley}, S.~K. {Reinhardt}, and C.~C.
  {Weems}, ``{Guided Region Prefetching: a Cooperative Hardware/Software
  Approach},'' in \emph{ISCA}, 2003.

\bibitem{hard}
P.~Zhou, R.~Teodorescu, and Y.~Zhou, ``{HARD: Hardware-Assisted Lockset-based
  Race Detection},'' in \emph{HPCA}, 2007.

\end{thebibliography}
\end{normalsize}
\end{document}